\documentclass[aps,preprint,showpacs,amsmath,amssymb,prb]{revtex4}
\usepackage{tabularx}

\usepackage{amsmath}

\usepackage[final]{graphicx}
\usepackage{dcolumn}
\usepackage{bm}
\def\extgra{pdf}

\usepackage{braket}

\usepackage{xcolor}
\renewcommand{\textcolor}[2]{#2}

\usepackage{verbatim}

\newcommand{\mm}[1]     {\ifmmode {#1} \else{}${#1}$\fi}
\newcommand{\mmm}[1]    {\ifmmode{}#1 \else{}${#1}$\fi}

\newcommand{\beq}[1]{\begin{equation}\label{#1}}
\newcommand{\eeq}{\end{equation}}

\def \cealge{\mm{\rm CeAlGe}}

\def\vec#1{{\mm{{\rm\bm{{\mathrm#1}}}}}}

\def\figsiz{\columnwidth}
\begin{document}
\def\figsiz{\columnwidth}
%
%
%


\title{On the magnetic contribution of itinerant electrons to neutron diffraction in the topological antiferromagnet \cealge}


\author{V.~Pomjakushin}
\affiliation{Laboratory for Neutron Scattering and Imaging LNS, PSI Center for Neutron and Muon Sciences, 5232 Villigen PSI, Switzerland~}

\author{A.~Podlesnyak}
\affiliation{Neutron Scattering Division, Oak Ridge National Laboratory, Oak Ridge, Tennessee 37831, USA}

\author{P.~Puphal}
\affiliation{Max Planck Institute for Solid State Research, Heisenbergstrasse 1, 70569 Stuttgart, Germany}

\author{S.~Shin}
\affiliation{Laboratory for Multiscale Materials Experiments LMX, PSI Center for Neutron and Muon Sciences, 5232 Villigen PSI, Switzerland}

\author{J.~S.~White}
\affiliation{Laboratory for Neutron Scattering and Imaging LNS, PSI Center for Neutron and Muon Sciences, 5232 Villigen PSI, Switzerland}

\author{E.~Pomjakushina}
\affiliation{Laboratory for Multiscale Materials Experiments LMX, PSI Center for Neutron and Muon Sciences, 5232 Villigen PSI, Switzerland~}
\date{\today}

\begin{abstract}


We report a neutron diffraction study of the magnetic structure of CeAlGe, a candidate topological semimetal that hosts a non-collinear, multi-$\vec{k}$ magnetic phase. By measuring both low- and high-momentum-transfer magnetic Bragg peaks within a single experimental setup, we refine a magnetic structure model based solely on localized Ce moments. This model, which differs from that obtained using only high-$Q$ data, quantitatively reproduces the observed intensities, including the $(000)$ zeroth-order magnetic satellites that are especially sensitive to subtle components of the modulation. While a contribution from itinerant electrons to the zeroth satellite cannot be definitively excluded, our analysis reveals no unambiguous evidence for such effects within experimental uncertainty.
The refined magnetic structures exhibit topologically nontrivial winding patterns, derived from the fitted magnetic parameters, that support localized, particle-like spin textures with half-integer topological charges. These features provide a natural microscopic origin for the observed topological Hall effect, establishing CeAlGe as a model system where magnetism and topology are intimately linked.

\end{abstract}

\pacs{75.30.Et, 61.12.Ld, 61.66.-f}

\maketitle

\section{Introduction}

Recently, we reported the discovery of topological magnetism in the candidate magnetic Weyl semimetal CeAlGe~\cite{cealge}. Using neutron scattering, we found that this system hosts an incommensurate, square-coordinated multi-\textbf{k} magnetic phase below $T_N$. The topological properties of a phase stable at intermediate magnetic fields parallel to the c-axis are suggested by the observation of a topological Hall effect (THE) and the integer topological charges $\mid Q\mid=1$ determined from the magnetic texture. The field- and pressure-dependence of THE, as well as its sensitivity to stoichiometry, were further investigated in Ref.~\cite{THE_Piva}.

The magnetic structure corresponds to the (3+2)-dimensional magnetic superspace group I4$_1$md1'(a,0,0)000s(0,a,0)0s0s, with Ce spins occupying the 4a Wyckoff position at $(0,0,z)$. 
The structure is based on the full star of the propagation vector {$[\pm k,0,0]$, $[0,\pm k,0]$}. Figure \ref{MS_exp} illustrates the magnetic structure in projection onto the (xy)-plane.

The long magnetic periodicity implies that itinerant electrons experience a slowly varying magnetic field as they move. This can cause conduction electrons to exhibit a spin polarization, which is responsible for THE, that oscillates with the same periodicity as the magnetic structure formed by the localized magnetic moments of Ce ions, thereby contributing to the magnetic scattering intensities. 
The general theory of neutron scattering from band electrons and a perfect electron fluid is presented in Sections 7.5 and 7.6 of Lovesey's book~\cite{LoveseyII}. The integrated intensity for this type of magnetic scattering depends on the scattering vector Q differently compared to the form factor of localized electronic shells, but overall scattering power appears to be comparable to that of the localized spins (see formula (7.112) in \cite{LoveseyII}). 
However, to our knowledge, no theoretical studies have quantitatively addressed how this scattering strength is partitioned between elastic and inelastic channels in antiferromagnetically ordered semimetals. If an elastic component of spatially modulated band-electron scattering does exist, its contribution is expected to appear  only in the so-called zeroth satellites, i.e., Bragg peaks such as $(\pm k, 0, 0)$ near $\vec{Q} = 0$, whereas the localized Ce moments contribute to both low- and high-$Q$ magnetic satellites.
%

In our previous study~\cite{cealge}, the zeroth satellites and the high-$Q$ magnetic satellites were measured on different instruments, which prevented a direct normalization of their intensities. To test for a possible Bragg contribution from itinerant electrons, it is essential to measure both the zeroth satellite and at least one additional high-$Q$ magnetic Bragg peak within the same experimental setup. This requires a momentum-transfer range of approximately $q \simeq 0.06$--$2$~\AA$^{-1}$, which goes beyond the capabilities of standard neutron diffraction configurations. Such measurements enable a direct intensity comparison with predictions from a magnetic model based solely on localized Ce moments.

In conventional metallic ferromagnets such as Fe, Co, and Ni, neutron scattering studies have extensively characterized the role of itinerant electrons in spin excitations~\cite{WINDSOR1977119,moriya1984itinerant}. In these systems, itinerant electrons contribute significantly to the dynamic magnetic response, leading to well-defined spin-wave dispersions and Stoner excitations. 
%
%
However, the role of itinerant electrons in neutron scattering from non-collinear, incommensurate antiferromagnets like CeAlGe remains less explored. In such systems, magnetic order originates primarily from localized moments, but the interaction between these moments and conduction electrons introduces additional complexity. As a result, direct analogies with ferromagnetic metals are of limited applicability, particularly for elastic neutron diffraction intensities, which are sensitive to long range spatial modulations rather than to dynamic excitations.

In this paper, we present neutron diffraction experiments in CeAlGe and analyze the possible contribution of itinerant electrons to the magnetic Bragg peaks. 
\textcolor{blue}{Our analysis suggests that the refined magnetic structures may support nontrivial topological charges, which could contribute to the topological Hall effect.}
%
%
Unfortunately, we cannot provide conclusive statements on the itinerant electron contribution due to the specifics of the magnetic structure factors in the studied Q-range. However, our study offers an interesting example where the potential role of conduction electrons in modulating magnetic diffraction intensities can be critically examined within a complex non-collinear magnetic structure.

\section{Samples synthesis and experimental details}
\label{exp}

%

Single crystals were grown using the traveling solvent floating zone method, as described in Refs.~\cite{Puphal2019,cealge}. For the powder diffraction experiments, the air-sensitive CeAlGe crystals were crushed and sealed in vanadium cans under helium atmosphere in a glovebox. We note that the samples used in the CNCS and DMC experiments described below originated from the same growth batch, but were not identical.

%
%
%
%

Magnetic neutron diffraction measurements were performed at the time-of-flight Cold Neutron Chopper Spectrometer (CNCS) \cite{CNCS,CNCS2} at the Spallation Neutron Source (SNS) at Oak Ridge National Laboratory, USA. Although CNCS is not designed as a diffractometer, its high flux, low background, and acceptable $Q$ resolution make it a highly suitable instrument for accurate analysis of elastic scattering. In particular, its capability to separate elastic scattering from inelastic signals enhances the signal-to-noise ratio and eliminates inelastic artifacts. 

Furthermore, the combination of low-angle detectors (starting at $3.5^\circ$) and a tunable wavelength range ($1.28$–$12.76$~\AA) enables diffraction measurements in the low-$Q$ region, which is conventionally considered small-angle neutron scattering (SANS) diffraction. The detectors, consisting of position-sensitive $^3$He tubes, cover scattering angles up to $135^\circ$ in the scattering plane. The overall detector background (electronic noise and background neutrons) amounts to less than $0.5$ counts per minute per meter of tube. To minimize beam divergence, a 60' solid-state collimator (SwissNeutronics AG, Switzerland) was placed upstream of the sample.

The data were collected using fixed incident neutron wavelengths of $4.96$, $7.26$, $10.78$, and $12.3$~\AA. A shorter wavelength is necessary to verify the nuclear and magnetic structure models, whereas a longer wavelength ($12.3$~\AA) provides unique experimental conditions, allowing the capture of both  zero-  and first-order magnetic satellites in the $Q$ range of 0.064–0.9~\AA$^{-1}$, which would otherwise not be accessible with a shorter wavelength.

For the crystal structure determination, we used the high-resolution diffractometer for thermal neutrons HRPT~\cite{hrpt} at the SINQ spallation source at the Paul Scherrer Institute (Switzerland), using wavelengths of 2.45~\AA{} and 1.89~\AA{}. In addition, for the refinement of the magnetic structure, we reanalyzed the powder diffraction dataset previously collected on the cold-neutron diffractometer DMC, as reported in Ref.~\cite{cealge}.

The determination of the crystal and magnetic structure parameters was performed using the {\tt FULLPROF}~\cite{Fullprof} program, utilizing its internal tables for neutron scattering lengths. The symmetry analysis was carried out using {\tt ISODISTORT} from the {\tt ISOTROPY} software~\cite{isod,isod2}, as well as various tools from the Bilbao Crystallographic Server, including {\tt MVISUALIZE}~\cite{Bilbao,ISI:000358484200010}.

\section{Crystal structure}

The crystal structure is well refined in the tetragonal space group $I4_1md$ (No. 109) using HRPT data, with an example of the diffraction pattern at $T = 10$~K and its Rietveld refinement shown in Fig.~\ref{nuc_10K_2p45A_HRPT}. All atoms occupy the $4a$ $(0,0,z)$ positions with $z = 0.58298(14)$, $0.16585(15)$, and $0$, and atomic displacement parameters (ADPs) $B = 0.228(41)$, $0.810(68)$, and $0.091(31)$ for Ce, Al, and Ge, respectively. The refined lattice parameters are $a = 4.26633(7)$~\AA{} and $c = 14.6606(2)$~\AA{}.

For the measurements performed at CNCS, we did not refine the structural parameters due to the limited $Q$-range. Only the crystal metric and overall scale factor were refined, as these are necessary for the subsequent magnetic fits. The quality of the fit is good and is illustrated in Fig.~\ref{nuc_10K_4p9A} for the diffraction pattern taken at $4.96$~\AA{}.

\section{Magnetic structure}
\label{mag_str}







The neutron diffraction intensities are dominated by very large nuclear peaks, which are located very close to the magnetic peaks due to the small propagation vector $\vec{k}$. For this reason, difference patterns - i.e., the difference between patterns measured at base temperature ($\simeq1.6$K) and in the paramagnetic state (10K) - were used to refine the magnetic structure. Such difference patterns contain purely magnetic scattering and are free from potential systematic uncertainties associated with the fitting of strong nuclear Bragg peaks, background subtraction, impurity phases, and other artifacts. 

The identification of the magnetic propagation vector was performed using the so-called le Bail refinement, in which all peak intensities are refined independently without assuming a structural model, allowing for an unbiased determination of the propagation vector $\vec{k}$. 
The propagation vector refined from the pattern measured with $\lambda = 4.96$~\AA{} was determined to be $\vec{k} = [g, 0, 0]$ with $g = 0.0743(2)$, corresponding to the SM-point of the Brillouin zone. Here, we follow the internationally established nomenclature for irreducible representation (irrep) labels and magnetic superspace groups (MSSG)~\cite{Bilbao,isod}.
The pattern collected with $\lambda = 12.3$~\AA{} does not contain any nuclear Bragg peaks, therefore, the previously determined value of $\bm{k}$ was used, with a slight adjustment of the wavelength to optimally fit the two observed magnetic Bragg peaks.
It is instructive to compare this propagation vector with the value refined from the DMC dataset, $k = 0.06597(15)$. The $\vec{k}$-vector exhibits a notable shift of approximately 12.6\% between the two experiments. This might be due to the fact that the samples and the measurement temperatures were not exactly the same.

\subsection{Magnetic Structure Refinements}
\label{mag_str_ref}

For completeness, we present the already published~\cite{cealge} magnetic model structure here in the notation of the magnetic superspace group (MSSG) $I4_1md1'(g,0,0)000s(0,g,0)0s0s$, generated by the mSM2 irrep and two arms of the k-vector star. There is a single Ce atom with only four (out of eight) nonzero parameters allowed by MSSG symmetry: sine (s) and cosine (c) amplitude modulations for k-vectors $\mathbf{k}=[g,0,0]$ (1) and $\mathbf{k}=[0,g,0]$ (2), along respective axes $(x,y,z)$. These are denoted as $\text{mxs1}=m_1$, $\text{mys2}=m_2$, $\text{mzc1}=m_3$, and $\text{mzc2}=m_4$, where the notation $(m_1, m_2, m_3, m_4)$ was used in our previous work~\cite{cealge}.  
The mcif file for this model is available in Supplementary Materials~\cite{SM} and as entry \#2.1.1 in the collection of magnetic structures with portable mcif-type files MAGNDATA~\cite{MAGNDATA}. For all models considered below, we use the same notation of parameters as $(m_1, m_2, m_3, m_4)$.

A comparison of neutron diffraction difference patterns (``1.6K - 10K'') measured on the DMC 
instrument at SINQ~\cite{cealge}  and on the CNCS instrument at SNS with incident wavelengths 
$\lambda = 4.506$ and 4.96~\AA\  (referred to as the P3 and P1 datasets, respectively) is presented in 
Fig.~\ref{dmc_vs_cncs}. Given that the primary goal is to verify the correspondence of the zero satellite intensity to the magnetic model assuming purely localized Ce moments, it is essential to compare the datasets from both instruments in more detail. 
The DMC dataset provides significantly better resolution, yielding sharper and well-defined Bragg peaks, which is essential for accurate magnetic structure refinement. The total effective counts, defined as $N_{\Sigma} = \sum (I/\sigma)^2$, where $I$ is the measured intensity and $\sigma$ is the associated error bar, are $3.9 \times 10^7$ for DMC and $1.7 \times 10^7$ for CNCS. The difference-based effective counts, given by $N_{d\Sigma} = \sum (I_1 - I_2)^2 / (\sigma_1^2 + \sigma_2^2)$, where $I_1$ and $I_2$ represent the measured intensities at base and 10K temperatures, respectively, are $1.5\times 10^4$ for DMC and $1.2 \times 10^4$ for CNCS. 
These values show that the statistical quality of the two datasets is comparable, allowing a direct comparison of the goodness of fits using $\chi^2$. 
Nevertheless, the DMC dataset is more suitable for high-precision refinements at high $Q$, due to its significantly better resolution.
Interestingly, despite its lower resolution, CNCS exhibits a lower and more uniform background, enhancing the signal-to-noise ratio (SNR), estimated as $\text{SNR} \propto {N_{d\Sigma}}/{\sqrt{N_{\Sigma}}}=2.9$, which is $\sim30$\% higher compared to DMC, implying that CNCS benefits from reduced background noise. This could be particularly important for detecting weak magnetic signals, especially in the CNCS dataset at $\lambda = 12.31$~\AA{}, which enables the observation of the zero-order magnetic satellite at very low $Q$ together with the first satellite.
For the CNCS dataset at $\lambda = 12.31$~\AA{}, the total effective counts are $N_{\Sigma} = 3.5 \times 10^5$, while the difference-based effective counts are $N_{d\Sigma} = 2.9 \times 10^3$, resulting in $\text{SNR} \approx 4.9$.

We have performed several types of fits. Model A, taken from our previous work~\cite{cealge} and described above, is based on the DMC dataset. This model uses fixed magnetic parameters for CNCS pattern. Table~\ref{mag_table} presents the refined values of the magnetic parameters and the goodness of fit $\chi^2$ for all models.  The fit quality for all models is illustrated in Fig.~\ref{dif_all_ABC}. 

Model B, refined from a single CNCS dataset at $\lambda=4.96$~\AA, covers a similar $Q$-range as the DMC diffraction pattern. The improvement of fit in comparison with Model A is marginal. A close inspection of the calculated profiles shows that Model B gives slightly different intensity for Bragg peaks (103), (110), and (112), which apparently modifies (makes larger in absolute values) the $z$-components of magnetization $m_3$ and $m_4$. These parameters remain in agreement within error bars, which are relatively large for the CNCS dataset, with the values from Model A. The key feature of the models A and B is that the $z$ components m1 and m2 are close in absolute values with opposite signs, which predicts significantly underestimated intensity of zero satellite, as one can see in Fig.~\ref{dif_all_ABC}. Model C is a combined fit of both CNCS datasets with $\lambda = 4.96, 12.31$~\AA. 

\section{Discussion}
\subsection{Diffraction results}
\label{discon}

Both models A and B, based on fits to the high-$Q$ patterns, predict significantly underestimated intensities for the zero satellites (those associated with the $(000)$ reflection), as shown in Fig.~\ref{dif_all_ABC}.  One possible interpretation of this discrepancy is that the missing intensity arises from contributions of itinerant electrons not accounted for in the localized-moment model. 

Model A, which provides the most reliable fit to the high-$Q$ diffraction data, being based on the best dataset (P3), predicts approximately four times lower intensity for the zeroth satellites compared to observations. As mentioned in the Introduction, there are currently no realistic theoretical calculations quantifying the elastic contribution of antiferromagnetically modulated band electrons. However, if we tentatively attribute this missing intensity to a contribution from itinerant electrons, we can proceed as follows.

The magnetic moment value in Model A varies between 0.12 and 1.16 $\mu_\mathrm{B}$. To explain the observed zeroth satellite intensity, we would need to approximately double the structure factor. According to formula (7.112) in Lovesey’s book~\cite{LoveseyII}, the integrated magnetic scattering intensity from a degenerate electron fluid $S(Q)$ (which we interpret here as an effective form factor) depends on the ratio $Q/k_\mathrm{F}$, where $k_\mathrm{F}$ is the Fermi wave vector.
ARPES measurements on CeAlGe~\cite{Belopolski} and band-structure calculations~\cite{Chang} suggest that $k_\mathrm{F}$ lies in the range of 0.05--0.2~\AA$^{-1}$. Within this range, the form factor at the zeroth satellite $S(Q)$ is expected to remain of order unity. This implies that the magnetic scattering amplitude from fully spin-polarized itinerant electrons could, in principle, be comparable to that of the localized Ce moments, on the order of 1~$\mu_\mathrm{B}$.
However, we emphasize that this interpretation remains highly speculative. The form factor used here is derived for an idealized free-electron gas, and to our knowledge, no theoretical framework exists that quantitatively describes how magnetic scattering from itinerant electrons in an antiferromagnetically ordered semimetal is partitioned between elastic and inelastic channels. Therefore, although such a contribution could phenomenologically account for the enhanced zeroth satellite intensity, it should be viewed as only a tentative working hypothesis rather than a quantitative estimation.

As shown in Table~\ref{mag_table}, Model C, which is based on a combined fit to the P1 and P2 datasets, provides an excellent description of the zeroth-order satellite in P2, with only a marginal increase in $\chi^2$ for the P1 dataset compared to Model B. Model B, like Model A for dataset P3, is based on a separate high-$Q$ fit.
In this context, Model C offers a consistent picture: it reproduces the diffraction data, including the zeroth-order satellites, without invoking a free-electron component whose neutron signal remains theoretically uncertain. However, this does not contradict the sizable topological Hall effect (THE) reported in Ref.~\cite{cealge}, which requires a real-space Berry curvature generated by non-coplanar spin textures arising from the same itinerant electron degrees of freedom we attempt to isolate via diffraction. This is because Models A and B, which lacks the zeroth-order intensity, still provides a similarly good fit to high-$Q$ data. Therefore, one can easily assume either a large or negligible band-electron contribution depending on the model, without significantly affecting the refinement quality.

This seemingly puzzling insensitivity of Models A and B to the intensity of the zeroth-order satellites finds its explanation in the structure of the magnetic structure factors $\vec{F}_m$, given by
$
\vec{F}_m(\vec{H}) = 
\sum_{j} \vec{S}_{\perp, j}^{(\vec{k})} \, e^{2\pi i\, (\vec{H} \cdot \vec{X}_j)}
$, 
where $\vec{H}=\vec{H}_0+\vec{k}$ is the satellite position in reciprocal space offset from the fundamental Bragg peak $\vec{H}_0$ for the respective $\vec{k}$-vector, which can be $\pm\vec{k}_x$ and $\pm\vec{k}_y$. The sum runs over the four Ce atoms at the positions $\vec{X}_j$ with the complex modulation amplitudes $\vec{S}_{\perp, j}^{(\vec{k})}$  perpendicular to the scattering vector $\vec{Q}$.  The amplitudes are constructed according to standard magnetic superspace group (MSSG) propagation formula from the sine and cosine components $m_1$, $m_2$, $m_3$, and $m_4$, described in Sec.\ref{mag_str_ref}, with experimentally determined values listed in Table~\ref{mag_table}.
The factor  $r_0 f({Q})/2$, which is product of $r_e\gamma/2$ and magnetic form factor is omitted in this formula for clarity. 

For the $(000)$ satellites, the magnetic structure factor reduces to $|\vec{F}_m| = m_3 + m_4$, depending solely on the sum of $m_3$ and $m_4$. Experimentally refined values for $m_3$ and $m_4$ from the high-$Q$ datasets P1 and P3 are close in absolute magnitude but opposite in sign, leading to a significant cancellation and resulting in the small calculated intensity of the $(000)$ satellites, as seen in Fig.~\ref{dif_all_ABC}.

There are 13 distinct magnetic satellites separated by $2\theta$, as one can see in Figs.~\ref{dif_all_ABC} and \ref{dif_DMC_ABC}. In practice, the number of contributing satellites is significantly higher due to powder diffraction multiplicity, with each reflection arising from 4 to 16 overlapping satellites. It is worth noting that the $(\vec{H}_0\pm \vec{k})$ satellites have different intensities, as observed experimentally, due to the complex phase factors associated with the magnetic structure. One can see that the models produce slightly different intensities for (101) and (004) satellites as shown in Fig.~\ref{dif_DMC_ABC_zoom}.

The $c$ lattice constant ($\simeq 15$~\AA) is about three times larger than the $a$ lattice constant, and within the available $2\theta$ range there are no pure $(h00)\pm\vec{k}_x$ satellites where the intensity would depend exclusively on $(m_3 + m_4)^2$, as it does for the $(000)$ satellites.
While components proportional to $(m_3 + m_4)^2$ are present in reflections such as $(00l)$, their contribution to the intensity is strongly suppressed.
This is because $\vec{S}_\perp$, the component of the magnetic moment perpendicular to the scattering vector $\vec{Q}$, introduces a geometrical factor.  In these peaks, the $m_3$, $m_4$ terms scale with $c^2 g^2$, while the $m_1$, $m_2$ terms scale with $a^2 l^2$. Given that the propagation vector value $g\simeq0.07$ is small, the $m_3$, $m_4$ terms acquire a suppression factor of about $20 l^2$ compared to the $m_1$, $m_2$ terms.

In mixed reflections such as $(h0l)$, the contributions from $m_3$ and $m_4$ enter either quadratically or mixed with $m_1$ and $m_2$, but no single reflection isolates the $m_3+m_4$ contribution cleanly for a direct test. Although visible contributions of $m_3$ and $m_4$ appear in peaks such as $(101)$ and $(103)\pm \vec{k}_x$, they are not decisive for directly validating the zero-satellite intensity predictions.
Naturally, $m_3$ and $m_4$ contribute to all magnetic reflections, and their values are obtained through refinement, as listed in Table~\ref{mag_table} along with their respective error bars. Figure~\ref{chi2} shows the dependence of the refinement quality (quantified by the $\chi^2$ value) on $m_3$ and $m_4$ for Model~A, based on the DMC dataset. The plot reveals a well-defined minimum at $m_3 = -0.21(4)$ and $m_4 = 0.32(5)$, corresponding to suppressed intensity of the $(000)$ satellites. A shift of $m_4$ to match the Model~C value ($m_4 = 0.47(1)$) clearly moves the system out of this minimum, illustrating the sensitivity of the fit to this parameter.
However, the absence of ``hallmark'' reflections dominated solely by the $(m_3 + m_4)$ combination precludes a straightforward cross-check with the $(000)$ satellite intensities. 

It is important to note that the $(000)$ satellite benefits from a distinct advantage: its intensity is strongly magnified by the Lorentz factor, $L \propto \lambda^3/(\sin \theta \sin 2\theta)$.
The ratio of the Lorentz factors between the $(000)$ satellite in the CNCS pattern P2 and the higher-angle peaks of pattern P1 (where $m_3$ and $m_4$ significantly contribute) at $2\theta > 60^\circ$ exceeds 500. Although pattern P1 contains a second low-angle peak, $(002)\pm \vec{k}$, whose uncorrected intensity—based on the structure factor and multiplicity—would be about five times that of the $(000)$ satellite, the Lorentz correction drastically enhances the $(000)$ intensity, as evident in Fig.~\ref{dif_all_ABC}. 
The effective statistics of pattern P2 containing two magnetic peaks (compared to thirteen in P1) is only three times lower than that of pattern P1 giving overwhelming preference to the $(000)$ satellite, effectively dictating the value of $(m_3 + m_4)$.

Under these circumstances, we cannot provide a conclusive answer to the question raised in the Introduction regarding  ``the itinerant contribution to the magnetic diffraction''. However, the combined analysis demonstrates that the localized-moment model provides a quantitatively accurate description of the observed $(000)$ satellite intensities, without requiring additional contributions beyond the ordered Ce moments.

\textcolor{blue}{
The uncertain itinerant contributions from the electrons could be experimentally verified by angle-resolved photoemission spectroscopy (ARPES), particularly by observing potential changes in the band structure across the magnetic transition. To our knowledge, no ARPES studies currently exist for CeAlGe. However, electrical resistivity measurements have shown a sharp anomaly at the N\'eel temperature~\cite{Hodovanets2018}, and the resistivity peak can be attributed to the gap opening at the Fermi surface. This suggests a Fermi surface reconstruction associated with the onset of magnetic order and supports the idea that itinerant electrons are coupled to the magnetic structure. Future ARPES experiments could test this scenario by directly probing the gap feature and determining whether its wave vector corresponds to the magnetic propagation vector.
Recent ARPES measurements on the isostructural compound CeAlSi~\cite{Cheng2024} demonstrate that the band structure and topological features such as Fermi arcs are highly sensitive to magnetism, highlighting the utility of this technique in rare-earth-based topological magnets.
Interestingly, temperature-dependent ARPES studies on the related compound PrAlGe~\cite{Forslund2025} showed that the bulk electronic structure remains largely unchanged across the ferromagnetic transition, suggesting that inversion symmetry breaking plays a dominant role in stabilizing topological features such as Fermi arcs. This finding contrasts with earlier neutron diffraction studies~\cite{Destraz2020}, which clearly established bulk magnetic order in the same compound, demonstrating an apparent insensitivity of the band structure to long-range magnetism - at least within the resolution of ARPES.
}

A useful point of comparison is the MnGe system, which is also metallic and exhibits a topological Hall effect~\cite{Pom2023}. Its magnetic unit cell is approximately two times smaller ($\simeq 30$~\AA), enabling observation of both the zero satellite and higher-$Q$ satellites within a single diffraction pattern using a wavelength of 2.45~\AA. In that case, the local-moment model similarly provided an excellent description of the data. Notably, even when the zero satellite was excluded from the refinement, its intensity, predicted using the refined parameters, agreed well with the experimental value. This differs from the present case, where such predictive consistency is absent.

\subsection{Topological Charges}

We now turn to the topological properties of the magnetic structure, which could potentially provide additional arguments in the selection of the appropriate magnetic model. In particular, the structure is expected to exhibit nontrivial topological features, such as the emergence of meron-like objects carrying half-integer topological charges, as previously  in Ref.~\cite{cealge}.

For an incommensurate structure, the size and direction of the atomic magnetic moments related by the propagation $\vec{k}$-vector are proportional to  $M\propto\cos(\vec{k}\cdot\vec{r}+\varphi)$, and in the limit of $\vec{k}\to0$  one can approximate the distribution of the magnetization density as spatially continuous because the difference in magnetization $\vec{M}$ between neighboring atoms vanishes. In the continuous limit of a magnetic structure propagating in the two in-plane directions ($xy$), the topological charge density is defined as the solid angle density formed by the normalized magnetization $\vec{n} = \vec{M}/|\vec{M}|$:
\beq{top_idx}
w(x,y)={1\over4\pi} (\vec{n} \cdot [\frac{\partial \vec{n}}{\partial x}\times \frac{\partial \vec{n}}{\partial y} ]),
\eeq
This expression is valid for any magnetic structure which propagates within the $ab$-plane and in particular for the double-$\vec{k}$ structure in CeAlGe. In this case, $w(x,y)$ describes the local topological winding (or charge) per unit area in the $xy$ plane. For magnetic textures with full 3D modulation, a different formulation is required to account for propagation along all spatial directions ($x$, $y$, and $z$) and their topological classification~\cite{Pom2023}.
Importantly, in the continuous limit, the winding density $w(x, y)$ is independent of both the absolute value of the propagation vector $k$ and the global phase offset $\varphi$. These parameters merely control the spatial scaling and global translation of the topological texture within the magnetic unit cell, without altering its intrinsic structure. We also note that, the global phase can be different for distinct propagation vectors such as the orthogonal $\vec{k}_1$ and $\vec{k}_2$ used in our case.
When this limit is valid, one can work in the continuous approximation neglecting the discrete nature of the lattice, and the topological charge can be simply computed using Eq.~\ref{top_idx}, as was done for MnGe~\cite{Pom2023}. However, the precise range of validity of this approximation remains unclear for real systems with finite values of the propagation vector~$k$.

In the general case of an incommensurate magnetic structure constrained by crystallographic symmetry, one encounters a fundamental difficulty in interpreting the topological charge distribution when magnetic moments are related by crystallographic rotations by of $\pi$, $2\pi/3$, $\pi/2$, or $\pi/3$. In such cases, the magnetic moments within the primitive cell (or zeroth cell) may differ by large angle close to the above crystallographic angles. 
Although propagation of the modulation to further unit cells may reduce the angular difference between neighboring moments, the periodicity of the structure ensures that large relative angles reappear regardless of how small the propagation vector $\vec{k}$ becomes. 
%

This imposes a fundamental limitation on treating such systems as continuous magnetization fields, as is often assumed in micromagnetic or Berry-phase-based models. In real crystals like CeAlGe, the magnetic structure is defined on a discrete lattice of Ce ions, and the exchange field experienced by itinerant electrons originates from localized spins modulated at long wavelengths but constrained by crystal symmetry. Nevertheless, one can numerically compute the local winding number density $w(x,y)$ based on the solid angle subtended by adjacent spins on the lattice, as shown below and originally presented in Ref.~\cite{cealge}. The physical interpretation of such calculations is that the spin of an itinerant electron effectively samples an averaged local exchange field produced by its nearest magnetic neighbors.


In CeAlGe, the Ce ions occupy $4a$ Wyckoff positions in the space group $I4_1md$, and their magnetic moments are related by symmetry operations, including a $4_1$ screw rotation along the $c$-axis. As a result, the moments at two Ce sites are not independent but are symmetry-related through the magnetic superspace group (MSSG) described in Sec.~\ref{mag_str_ref}. The magnetic moments are generated by the formula:
\begin{align}
\vec{M}_1(x, y) &= 
m_1 \sin(\tilde{k} x)\, \hat{x} + 
m_2 \sin(\tilde{k} y)\, \hat{y} + 
\left[ m_3 \cos(\tilde{k} x) + m_4 \cos(\tilde{k} y) \right] \hat{z} \notag \\
\vec{M}_2(x, y) &= 
m_2 \sin(\tilde{k} x)\, \hat{x} + 
m_1 \sin(\tilde{k} y)\, \hat{y} + 
\left[ m_4 \cos(\tilde{k} x) + m_3 \cos(\tilde{k} y) \right] \hat{z}, 
\label{m1m2}
\end{align}
where $x$ and $y$ are fractional coordinates and  $\tilde{k} = 2\pi k$ with  $k$ denoting the size of the propagation vectors $\vec{k}_x = (k, 0, 0)$ and $\vec{k}_y = (0, k, 0)$. These moment expressions reflect the enforced symmetry relations between Ce sites due to the $4_1$ screw axis and the double-$\vec{k}$ modulation and illustrate the complex interplay between in-plane and out-of-plane modulations.
The continuous limit is only meaningful for this magnetic symmetry when $m_1 = m_2$, ensuring that swapping the $x$ and $y$ components does not transform a vanishing local solid angle, arising solely from the propagation formula, into a finite one between Ce1 and Ce2 spins. Likewise, $m_3 = m_4$ is required so that the $z$-component modulations propagate equally for both Ce atoms along the $x$ and $y$ directions. These constraints are necessary for the topological winding density $w(x, y)$ to remain smooth and well-defined in the continuous limit. When these conditions are approximately satisfied, topological properties can be readily calculated using (\ref{top_idx}), because they should not depend on the discrete sampling of the magnetization on the atomic lattice.

In the present experimental case, however, the conditions for the continuous approximation are not met, and the winding density must instead be computed directly on the discrete lattice, an approach that introduces additional complications, as discussed below.
When viewed along the $z$-axis, the Ce atoms on the two crystallographic sites form square plaquettes in projection onto the tetragonal ($ab$) plane, as illustrated in Fig.~\ref{MS_exp}, where Ce1 and Ce2 atoms are shown by small green and large blue circles, respectively. The solid angle $\Omega$ subtended by the spins on two adjacent triangles forming a plaquette defines the local topological charge density, $w = \Omega / 4\pi$. The solid angle for each triangle is computed using the Van Oosterom-Strackee formula~\cite{van1983solid}. For completeness we present the in-plane coordinates $(x, y)$ for $\vec{M}_1$ (Ce1) and $\vec{M}_2$ (Ce2): Ce1 if $x, y$ are both integers or both half-integers; Ce2 if one is integer and the other half-integer. This follows from the symmetry of the $4a (0,0,z)$ Wyckoff position in space group $I4_1md$.

\begin{figure}
  \begin{center}

    \includegraphics[width=\count1cm]{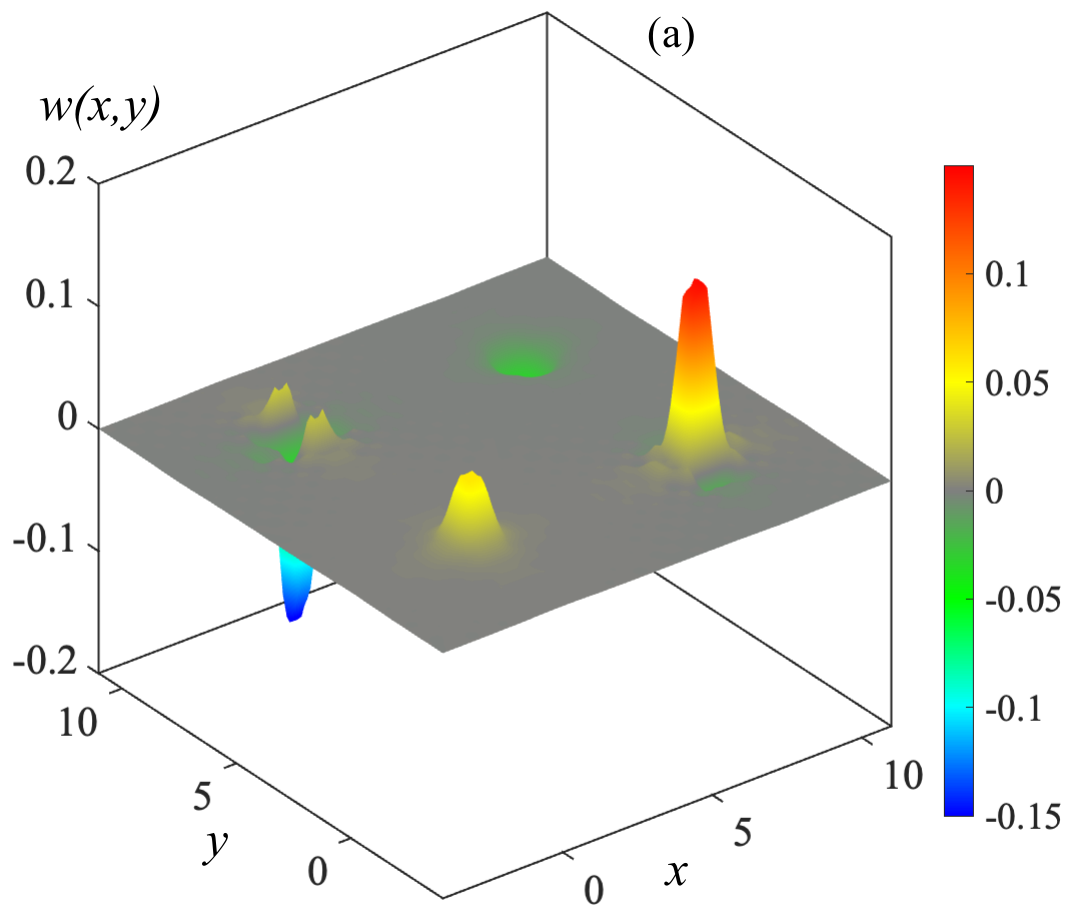} 
    \includegraphics[width=\count1cm]{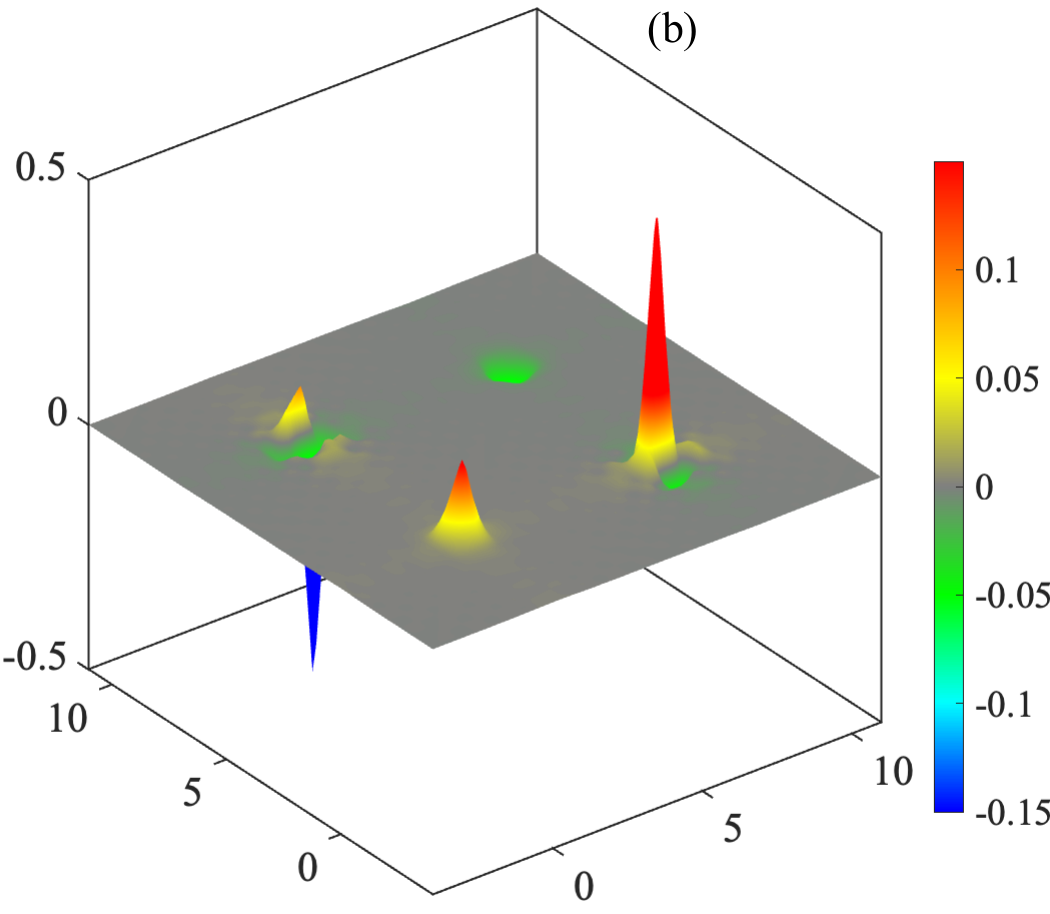} 
    \includegraphics[width=\count1cm]{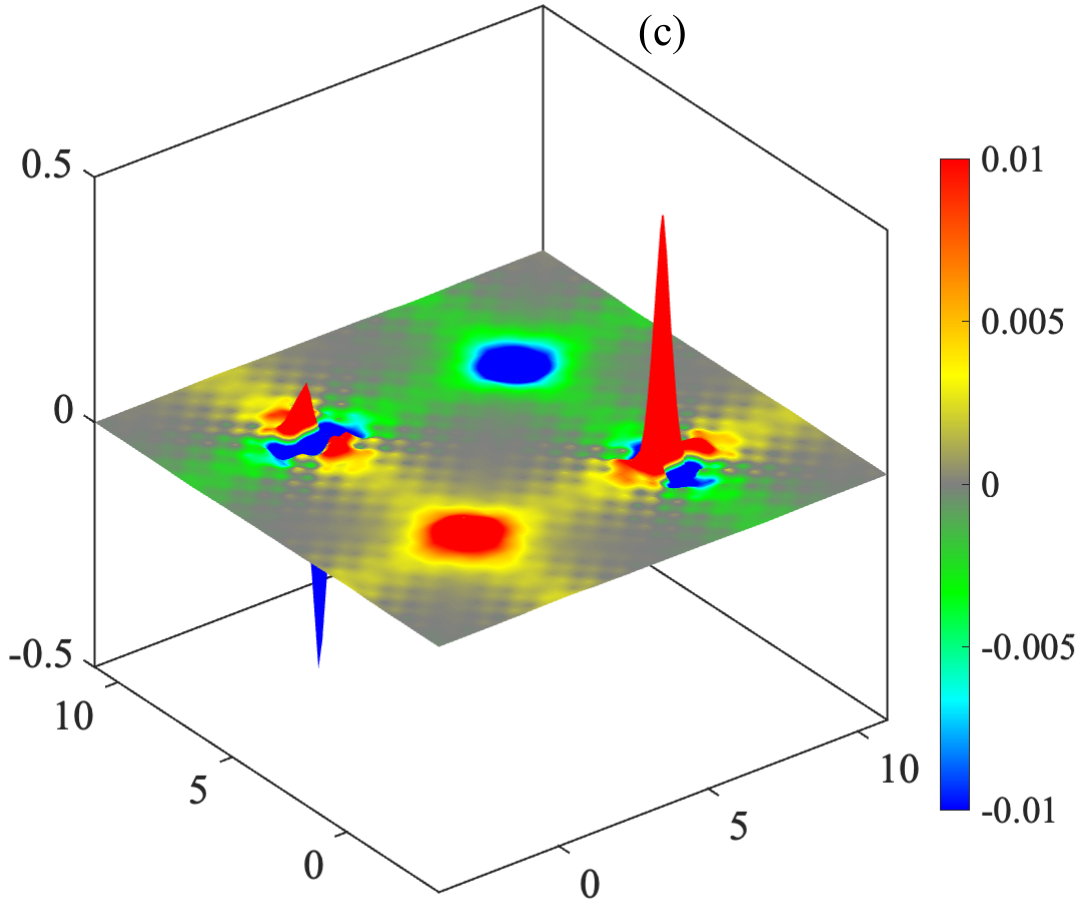} 
    \includegraphics[width=\count1cm]{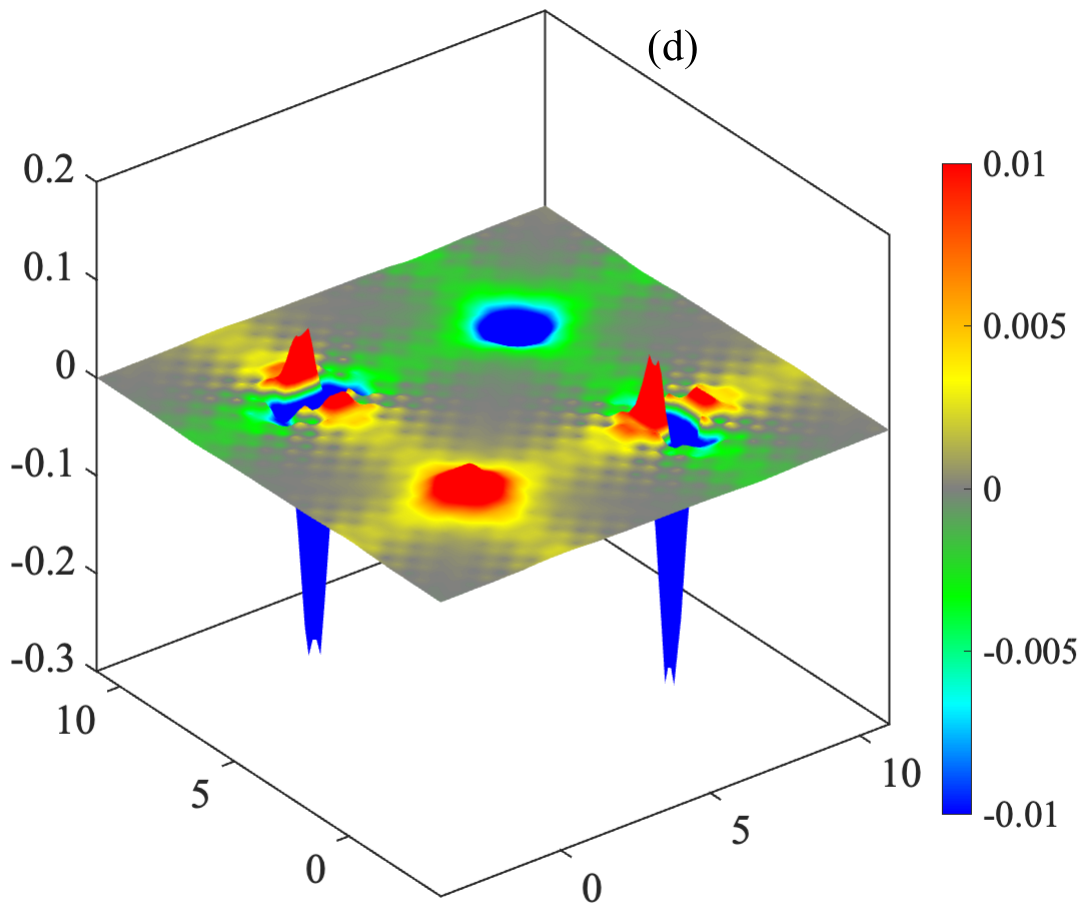} 

    
  \end{center}
\caption{
Topological charge density $w(x,y)$ calculated for the refined magnetic structures, where $x$ and $y$ are fractional coordinates of the plaquette in lattice units. Approximately one magnetic unit cell is shown. Top row: (a) Model A (left), refined from DMC data with $k \simeq -0.0665$. (b) Model B (right) with an overall phase $\varphi = 0.05$. Bottom row: (c) Model C with $\varphi = 0.05$ (left) and (d)  $\varphi = 0$ (right). Models B and C were refined from CNCS data with $k \simeq -0.0743$. Refined parameters for all models are listed in Table~\ref{mag_table}. Positive and negative topological charges appear as localized features in $w(x, y)$, concentrated near the quadrant centers of the magnetic unit cell. Each quadrant carries a charge of $\pm 1/2$.
}
  \label{topo}
\end{figure}

The spatial distribution of $w(x, y)$ is shown in Fig.~\ref{topo}. One observes regions of positive and negative winding density localized near the centers of the four quadrants, specifically at positions $(0,0)$, $(0, T/2)$, $(T/2, 0)$, and $(T/2, T/2)$, where $T = 1/k$ is the period of the magnetic modulation. The corresponding values of $T$ are 13.495(35) and 15.16(3) lattice units for patterns P1 (CNCS) and P3 (DMC), respectively. These quadrant centers represent the core regions of the topological spin textures and coincide with local minima in the magnitude of the magnetization $|\vec{M}|$.

As shown in Ref.~\cite{cealge}, a key feature of the CeAlGe magnetic structure is the presence of particle-like textures carrying topological charges $Q = \pm 1/2$, defined as integrals of $w(x, y)$ over localized areas of the magnetic unit cell.  When a magnetic field is applied along the $z$-axis, the system enters a topological phase characterized by a nonzero topological Hall effect (THE), driven by the emergence of a finite net topological charge per magnetic unit cell within an intermediate field range.

One can see that models A and B both produce sharp extrema in the centers of the four quadrants. The sum of the $w(x,y)$ over each quadrant amounts to $\pm1/2$.
The $z$-components of the magnetic moments play a central role in shaping the sharpness of the extrema in $w(x, y)$, with the most localized features occurring near the condition $m_3 = -m_4$. As illustrated in Fig.~\ref{topo}, Model C yields significantly lower extremum intensities compared to Model A, as the absolute values of $m_1$ and $m_2$ diverge in Model C, to match the intensity of zeroth satellites as discussed above.  In particular, the peak localized at $(0, 0)$ is weaker than the one at $(T, 0)$ only by a factor 2 in Model A, but about 10 in Model C.   Nevertheless, the integral of $w(x, y)$ over each quadrant remains $\pm 1/2$, and thus the net topological structure remains unchanged. From this perspective, the essential topological properties, including the presence of THE, are preserved and consistent with our earlier findings.

A difficulty here in the analysis is that the density calculated using (\ref{m1m2}) on the discreet lattice depends on both $k$-vector value and overall phase $\varphi$. This however does not have effect on the topological properties, which should not be dependent on these parameters. 
An incommensurate magnetic structure within a crystal naturally includes all possible magnetic unit cells shifted by a phase due to the aperiodicity of the modulation. Only in special commensurate cases when the magnetic period $T = 1/k$ is an integer all magnetic cells are with the same phase. We note that this phase dependence is intrinsic to any incommensurate case, including models that admit a continuous limit (as illustrated in Figs.S2 and S3 of the Supplementary Materials\cite{SM}), which, however, can be treated within the continuum approximation by directly using Eq.~(\ref{top_idx}).
 
\begin{figure}
  \begin{center}

    \includegraphics[width=\count1cm]{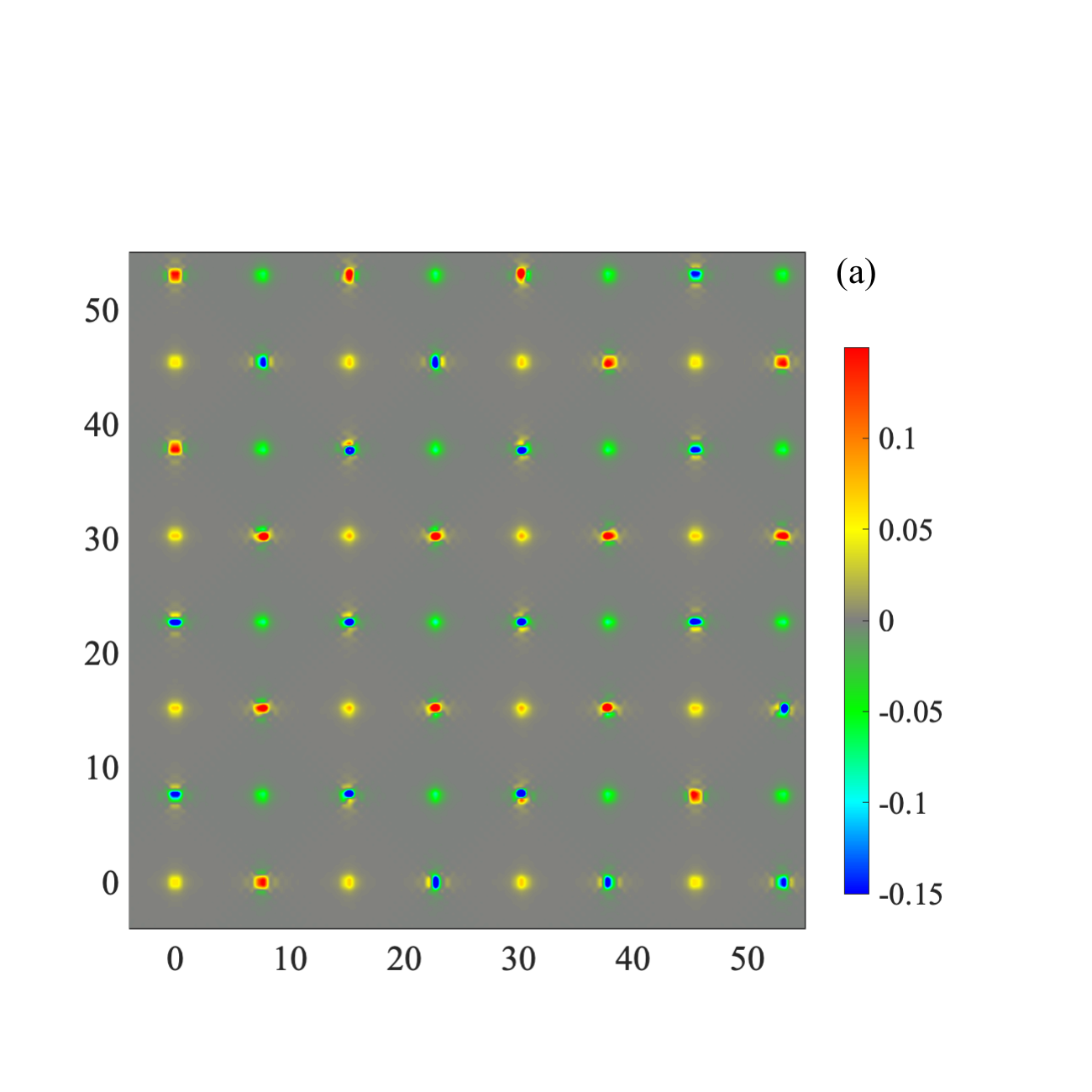} 
    \includegraphics[width=\count1cm]{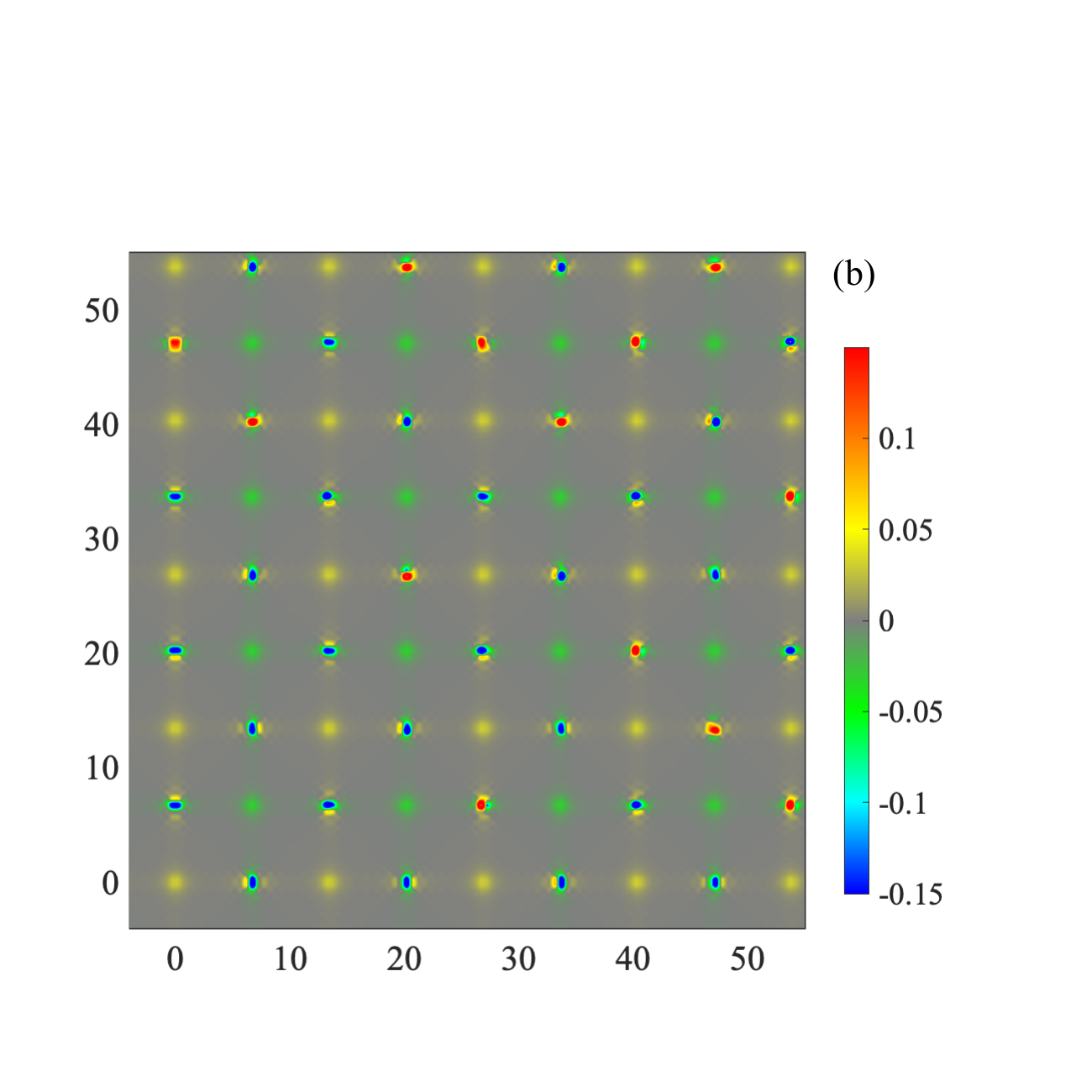} 
    
  \end{center}
\caption{
Topological charge density $w(x,y)$ calculated for Model A (left (a)) and Model C (right (b)), where $x$ and $y$ are fractional plaquette coordinates in lattice units. The magnetic periods: 15.16(3) and 13.495(35) crystallographic lattice units, respectively. Approximately $4\times4$ magnetic unit cells are shown. Four positive (red) and negative (blue) topological charges appear as localized features in $w(x, y)$, concentrated near the quadrant centers of each magnetic unit cell. The color scale is saturated to enhance visibility of these features, each carrying a charge $Q \simeq \pm 1/2$ when integrated over the quadrant. The total topological charge summed over the entire lattice is zero.
}

  \label{10x10}
\end{figure}

This behavior is illustrated in Fig.~\ref{topo}, which shows the spatial distribution of $w(x, y)$ for representative values of $\varphi$ in Model C. For example, a phase $\varphi = 0$ results in a net topological charge $Q = -1$ within the magnetic unit cell, while $\varphi = 0.05$~rad yields $Q = 0$. At $\varphi \simeq 0.12$, the sharp minima located at quadrants $(0,T/2)$ and $(T/2,0)$ switch to sharp maxima (as shown in Fig.S1 of the Supplementary Materials\cite{SM}), giving a net charge of $Q = +1$. This variation reflects how $Q$ depends on the choice of the magnetic cell of magnetic modulation, and corresponds to shifting the magnetic unit cell relative to the origin  given by (\ref{m1m2}). This property might be in general overlooked and to explain this in more details we present the Fig.~\ref{10x10} where several magnetic unit cells are shown. 
One can see that the net topological charge per magnetic unit cell, summed over the four quadrant centers, visible as bright red and blue spots, can switch between $Q = -1, 0, 1$  depending on the global phase, i.e., the relative shift of the magnetic cell with respect to the crystal lattice. This phase dependence is an intrinsic feature of incommensurate modulations defined on a discrete atomic lattice and does not indicate any physical singularity or discontinuity. 

For example, in Model C, the magnetic period is $T = 13.495(35)$, which can be well approximated within experimental uncertainty by the rational fraction $k = {14}/{188}$. This corresponds to a magnetic modulation that spans 188 crystallographic unit cells over 14 magnetic periods. As a result, the spatial pattern of the topological charge density $w(x, y)$ approximately repeats every 188 unit cells. 

\begin{figure}
  \begin{center}
    \includegraphics[width=\count1cm]{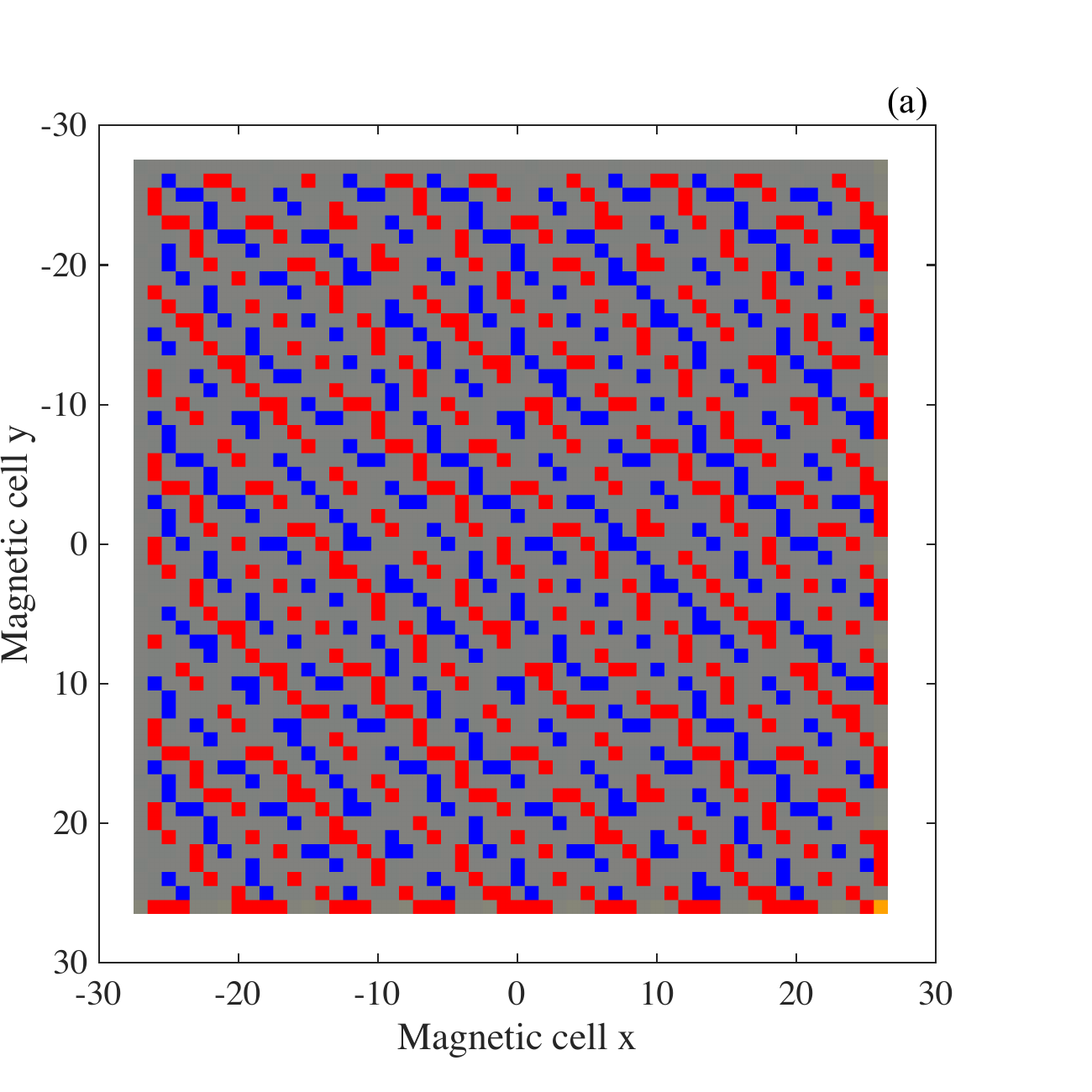} 
    \includegraphics[width=\count1cm]{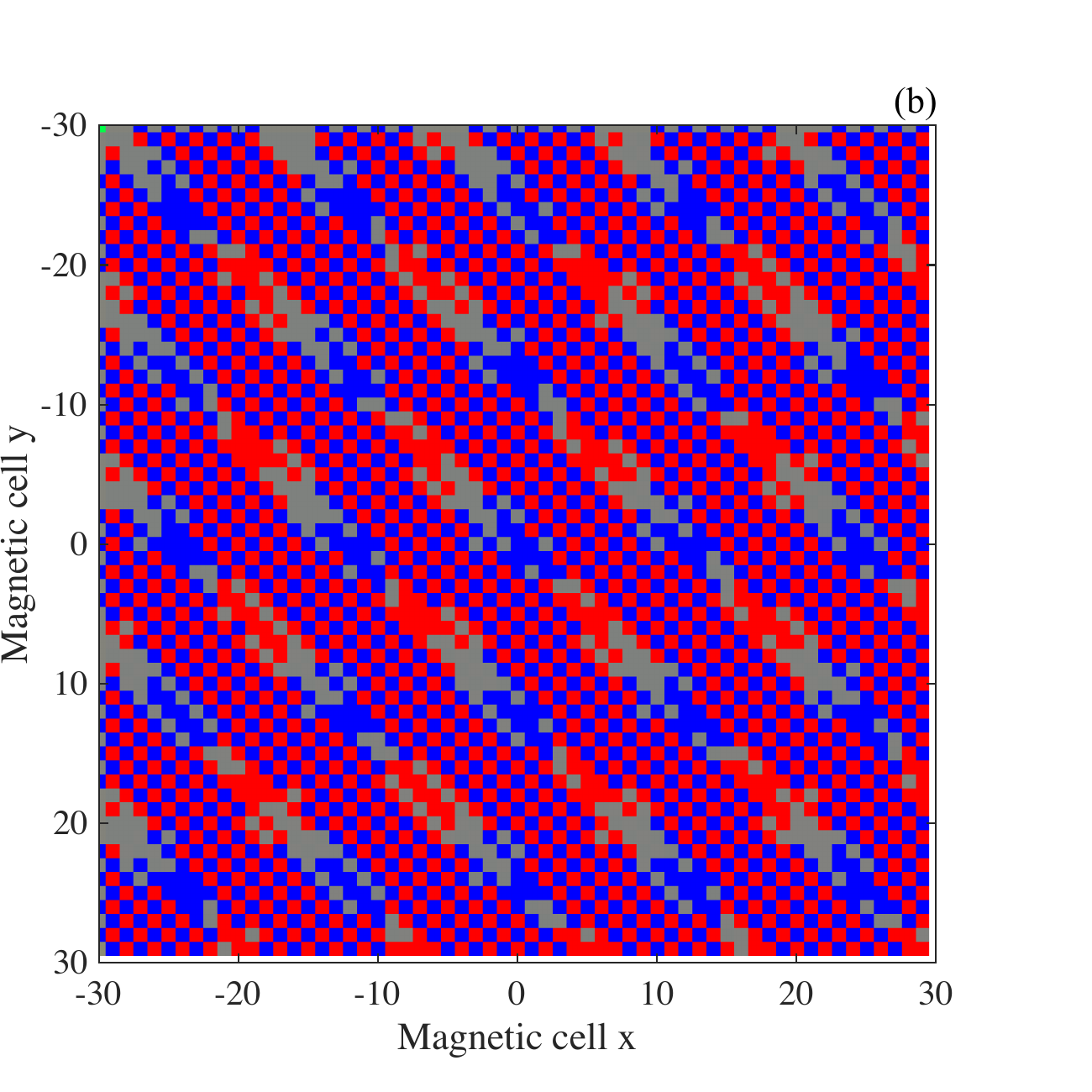} 
    \includegraphics[width=\count1cm]{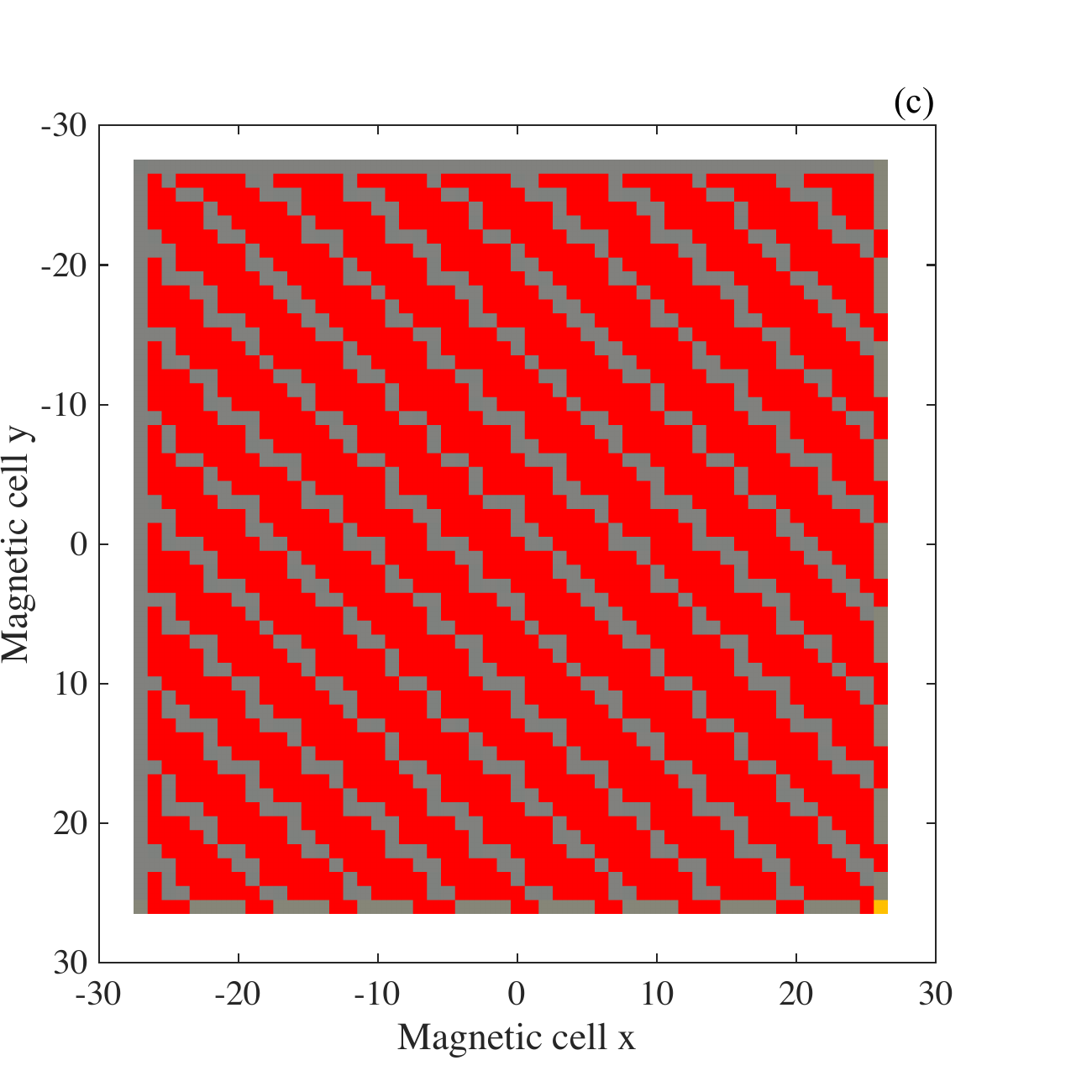} 
    \includegraphics[width=\count1cm]{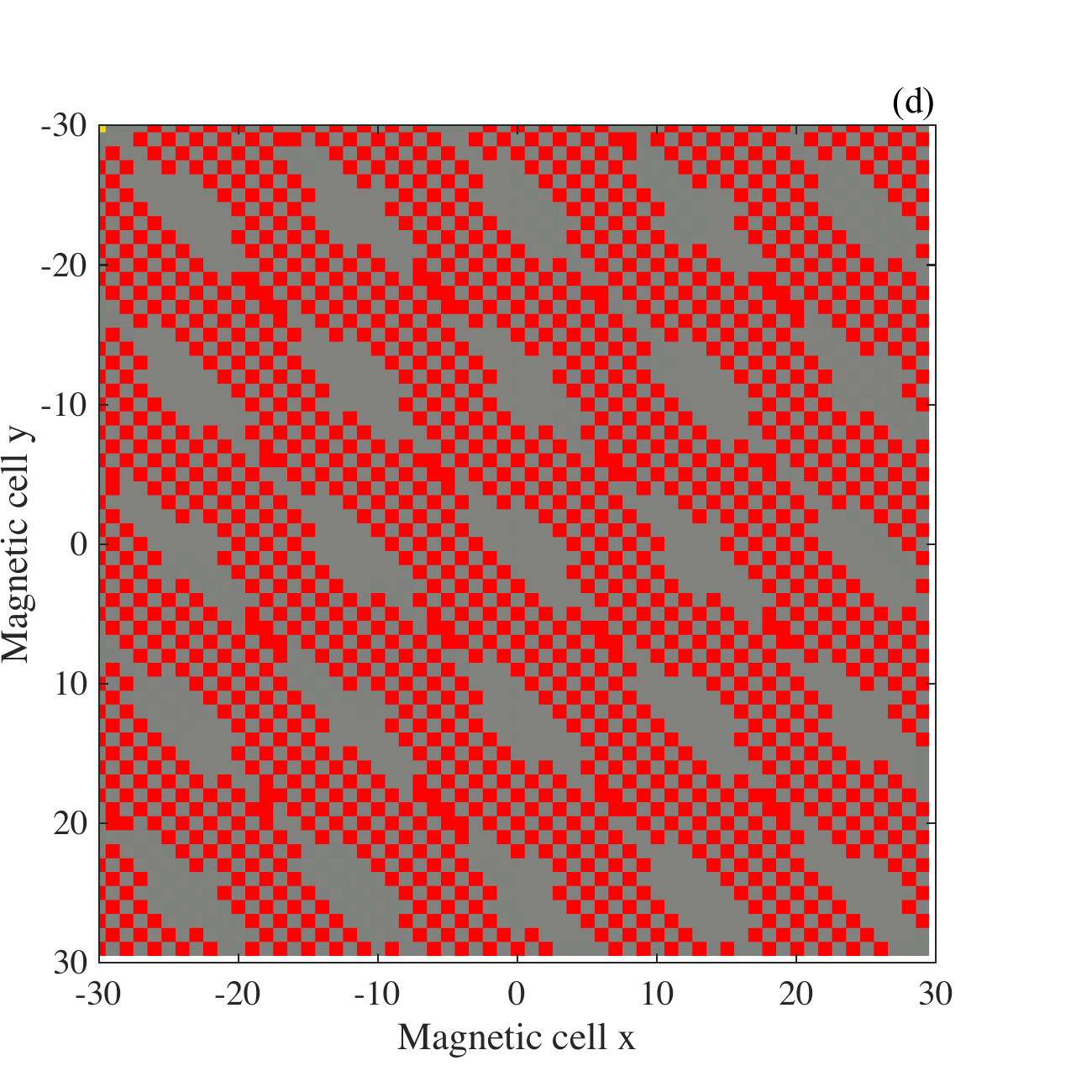} 

  \end{center}
\caption{
Topological charge $Q$ per magnetic unit cell calculated for Model A (left (a)) and Model C (right (b)). The top row corresponds to the zero-field structures with parameters listed in Table~\ref{mag_table}, while the bottom row includes an additional ferromagnetic component along the $c$-axis for models A (c) and C (d). The axes represent integer indices $(x, y)$ of the magnetic unit cells, defined in units of the magnetic periods: 15.16(3) and 13.495(35) crystallographic lattice units, respectively. Each magnetic cell carries a topological charge $+1$, $0$, or $-1$, shown in red, gray, and blue, respectively. See text for discussion of the topological implications and statistical measures. 
}

  \label{Qbycell}
\end{figure}

The physically relevant quantity is the average topological charge per magnetic unit cell across the entire sample, which vanishes at zero field: $\langle Q \rangle = 0$. Figure~\ref{Qbycell} shows the topological charge $Q$ computed on a cell-by-cell basis for the magnetic unit cells. The origin of the zeroth cell was chosen at $(-4, -4)$ in crystal lattice units to ensure that the topological core is not located at the boundary of a magnetic unit cell (see Fig.~\ref{10x10}). Despite large visual differences in the spatial distribution of $Q$, both models are topologically equivalent: the net averages are $\langle Q \rangle = -0.002(10)$ for Model A and $-0.02(2)$ for Model C, while the mean deviation from integer-quantized values ($-1$, $0$, $+1$) is 0.0025 and 0.0008, respectively, confirming the stability and precision of the calculations, which were averaged over approximately 2600 magnetic unit cells.

\textcolor{blue}{We note that the global phases associated with the two propagation vectors are not fixed by neutron diffraction, and different choices correspond to rigid real-space translations of the magnetic texture along the $x$- and $y$-axes by $\Delta x, \Delta y = \varphi_{x,y}/(2\pi k)$. However, such shifts do not affect the magnetic topology or the energy of the system. This invariance is a natural consequence of the aperiodic character of incommensurate magnetic structures.}

In both Models A and C, the average topological charge $\langle Q \rangle$ becomes nonzero upon applying a magnetic field, modeled by introducing a uniform ferromagnetic component $m_f$ along the $c$-axis, as previously discussed in Ref.~\cite{cealge}. This artificial field-induced configuration can be implemented in two ways: by directly adding a constant $m_f$ term to the $z$-component of the magnetization, or by adding it while renormalizing the total spin size to preserve a fixed magnitude, effectively mimicking canting. 
Although the overall behavior is the same in both cases, we present the latter method, which appears to be more physically relevant.

The spatial distribution of the topological charge $Q$ per magnetic unit cell for representative values of $m_f = 0.14$ and $0.26$ is illustrated in Fig.~\ref{Qbycell} (bottom row), for Models A and C, respectively. At these points, the net charge is near its maximum, with $\langle Q \rangle = 0.77(1)$ for Model A and $\langle Q \rangle = 0.88(2)$ for Model C. 
While not all magnetic cells acquire the same nonzero charge, the majority exhibit a dominant quantized value of $+1$. The remaining cells carry charges of $0$ or the opposite sign, maintaining the overall quantized nature of $Q$. This redistribution of topological charge in the presence of finite $m_f$ indicates a transition into a topological phase with a broken balance between merons and antimerons, consistent with the observed topological Hall effect (THE).

%
The dependence of $\langle Q \rangle$ on $m_f$ is summarized in Fig.~\ref{Qofmf}, which reveals a broad intermediate-field region where the net topological charge per magnetic unit cell becomes significantly nonzero. This signals the emergence of a topologically nontrivial phase induced by the finite out-of-plane spin component. The field range associated with this regime is similar in both models, but the precise values and sign of $\langle Q \rangle$ differ due to differences in their z-component parameters (Table~\ref{mag_table}).

As $m_f$ increases further, the average topological charge eventually drops back to zero above a threshold field $m_\mathrm{cr}$. This upper transition does not mark a crossover into a fully ferromagnetic state. Instead, the spin configuration remains a noncollinear, canted antiferromagnetic. 
Although the $z$-component of magnetization is still \textcolor{blue}{modulated but positive}, and the in-plane AFM modulations persist, the local winding density $w(x,y)$ becomes small, though not strictly zero, with the zero total charge per magnetic cell across the whole lattice.
This behavior contrasts with the zero-field ($m_f = 0$) case, where local topological textures—merons and antimerons—are present and balanced such that $\langle Q \rangle = 0$ globally. In the high-field regime ($m_f > m_\mathrm{cr}$), the winding disappears and all cells become topologically trivial. 
This transition thus marks the exit from the topological phase into a modulated but topologically inert magnetic state. The suppression of winding in this regime aligns with the disappearance of the topological Hall effect observed experimentally at higher fields~\cite{cealge}.

These results demonstrate that both models exhibit topological features at intermediate fields and are, in principle, compatible with the observed topological Hall effect (THE). Therefore, topological arguments alone cannot be used to discriminate between the models. However, as discussed above, Model A would require a substantial itinerant electron contribution to account for the zeroth satellite intensity, while Model C provides a satisfactory description based solely on localized Ce moments.

\section{Conclusions}
\label{conclusion}

We have investigated the magnetic structure of CeAlGe using neutron diffraction over a wide momentum-transfer range. The data are well described by a  multi-$\vec{k}$ magnetic structure based on localized Ce moments. Both high-$Q$ magnetic satellites and the zeroth-order $(000)$ satellites - potentially sensitive to itinerant electron contributions - are quantitatively captured within this model.
Including the zeroth satellites in the refinement yields a modified magnetic structure compared to high-$Q$-only fits, particularly affecting the distribution of moment components along the $z$-axis (i.e., out of the propagation plane). Nevertheless, no additional contributions beyond the ordered Ce moments are required to explain the observed intensity pattern.
The possibility of itinerant contributions cannot be entirely excluded, however, as magnetic structure factors at high $Q$ are relatively insensitive to the $z$-modulated components of the spin structure - precisely those that dictate the intensity of the zeroth-order satellites. In the absence of magnetic reflections that selectively probe these specific Fourier components, a direct test of the itinerant scenario remains inconclusive. Thus, while our findings strongly support a localized moment model, they do not definitively rule out a contribution from itinerant magnetism.
Nevertheless, this study provides an instructive example of how conduction electron effects might manifest in neutron diffraction on complex incommensurate magnetic structures. 

The refined magnetic structure in CeAlGe supports localized spin textures propagating in tetragonal $ab$-plane that subtend finite solid angles, consistent with half-integer topological charges. This provides a natural explanation for the observed topological Hall effect and highlights the intrinsic link between magnetism and topology in this material.
The local topological charge density $w(x, y)$, computed from spin configurations on the discrete Ce lattice, depends on the propagation vector and global phase due to the incommensurate nature of the magnetic modulation. Consequently, the topological charge per magnetic unit cell can vary between $Q = -1$, $0$, and $+1$, depending on the cell’s position. However, in a real crystal, the system effectively samples all phase offsets, resulting in a zero net charge $\langle Q \rangle = 0$ at zero field.
When a uniform magnetic field is modeled along the $c$-axis, the system enters a topological phase with a nonzero average charge per magnetic unit cell. This topological signature persists over a broad field range before vanishing again at higher fields, where the spin structure remains modulated but becomes topologically trivial. These results confirm that topological features in CeAlGe are both robust and tunable by magnetic field.

\section*{Acknowledgments}

This research utilized resources at the Spallation Neutron Source, a DOE Office of Science User Facility operated by Oak Ridge National Laboratory. The beam time was allocated to CNCS on proposal number IPTS-24824. S. S. acknowledges support from the SNSF under Grant No. 188706.
Part of this work was performed at the Swiss Spallation Neutron Source (SINQ), Paul Scherrer Institut (PSI), Villigen, Switzerland.

We gratefully acknowledge valuable contributions from Naoya Kanazawa, Victor Ukleev, Dariusz J. Gawryluk, Junzhang Ma, Muntaser Naamneh, Nicholas C. Plumb, Lukas Keller, and Robert Cubitt, as recognized in our previous study~\cite{cealge}.

The CNCS data that support the findings of this paper are openly available \cite{CNCSdata}.

 \bibliographystyle{unsrturl} 
\bibliography{CeAlGe_itinerant2arx_arx.bbl}


\clearpage

\begin{table}

\caption{
Refined magnetic amplitudes $(m_1, m_2, m_3, m_4)$ in $\mu_\mathrm{B}$ and $\chi^2$ values for different models. Minimum and maximum values of the magnetic moments (also in $\mu_\mathrm{B}$) are listed for comparison.
Model A, taken from our previous work~\cite{cealge}, is based on the DMC dataset measured with $\lambda = 4.506$~\AA{} (denoted as P3). 
Models refined from CNCS datasets at $\lambda = 4.96$~\AA{} (P1) and $\lambda = 12.31$~\AA{} (P2) include: Model B, based on P1 only, and Model C, a combined refinement using both CNCS datasets. 
The goodness-of-fit values $\chi^2$ are derived from standard Rietveld refinement reliability factors~\cite{Fullprof}, with corrections for background and evaluated only over regions containing Bragg peaks.
}


\label{mag_table}
\begin{center}
%
%
%

\begin{tabular}{l|l|l|l|l|l}
\hline
Model & $m_1$ & $m_2$ & $m_3$ & $m_4$ & $\chi^2$ \\ \hline
A & 0.457(9)  & 1.07(1)  & -0.21(4)  & 0.33(5)  & 1.88 (P3) DMC \\
\multicolumn{5}{r|}{min/max $x,y$: $\pm$1.07 , $z$: $\pm$0.54 } & 2.81 (P1) CNCS \\
\multicolumn{5}{r|}{min/max $|m|$: 0.12, 1.16} & 4.83 (P2) CNCS \\
\hline
B & 0.42(1)  & 1.06(1)  & -0.30(5)  & 0.37(7)  & 2.37 (P1) CNCS \\ 
\multicolumn{5}{r|}{min/max $x,y$: $\pm$1.05 , $z$: $\pm$0.67 } & 6.48 (P2) CNCS \\ 
\multicolumn{5}{r|}{min/max $|m|$: 0.07, 1.14} & 1.99 (P3) DMC \\ 
\hline 
C & 0.41(1)  & 1.042(10)  & -0.21(1)  & 0.47(1) & 2.43 (P1) CNCS \\ 
\multicolumn{5}{r|}{min/max $x,y$: $\pm$1.04 , $z$: $\pm$0.68} & 0.866 (P2) CNCS \\ 
\multicolumn{5}{r|}{min/max $|m|$: 0.25, 1.12} & 2.01 (P3) DMC \\ 
\hline
\end{tabular}

\end{center}
\end{table}


\def\extgra{pdf}
\def\figsiz{\textwidth}
\def\figsiz{8cm}
\def\figsizcm{8}

\def\figsiz{\textwidth}

\begin{figure}
  \begin{center}
    \includegraphics[width=\figsiz]{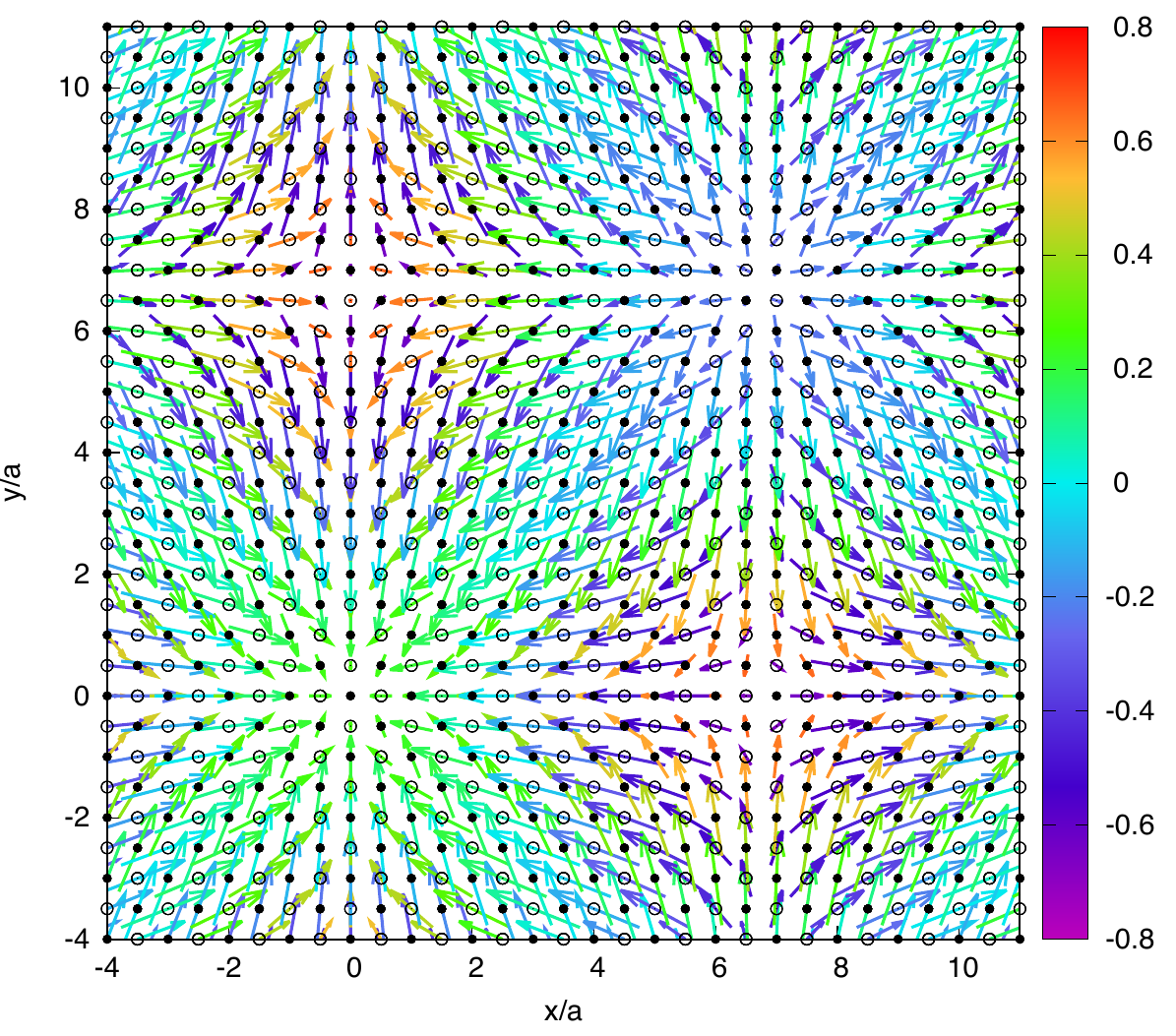}    
    
  \end{center}

\caption{
Magnetic structure of CeAlGe corresponding to the experimentally refined Model C (see Table~\ref{mag_table}). 
\textcolor{blue}{The $x$- and $y$-axes are given in units of the crystallographic unit cell. 
The magnetic modulation has a period of approximately 15 unit cells; the figure displays one full magnetic period (15~$\times$~15 cells) projected onto the $xy$-plane. 
Ce1 and Ce2 sites are represented by small filled and large open black circles, respectively. 
The out-of-plane ($z$) component of the magnetic moments is indicated by color. 
The moments form ferromagnetic chains along the $z$-axis at each $(x,y)$ position.}
}

\label{MS_exp}
\end{figure}

\begin{figure}
  \begin{center}
    \includegraphics[width=\figsiz]{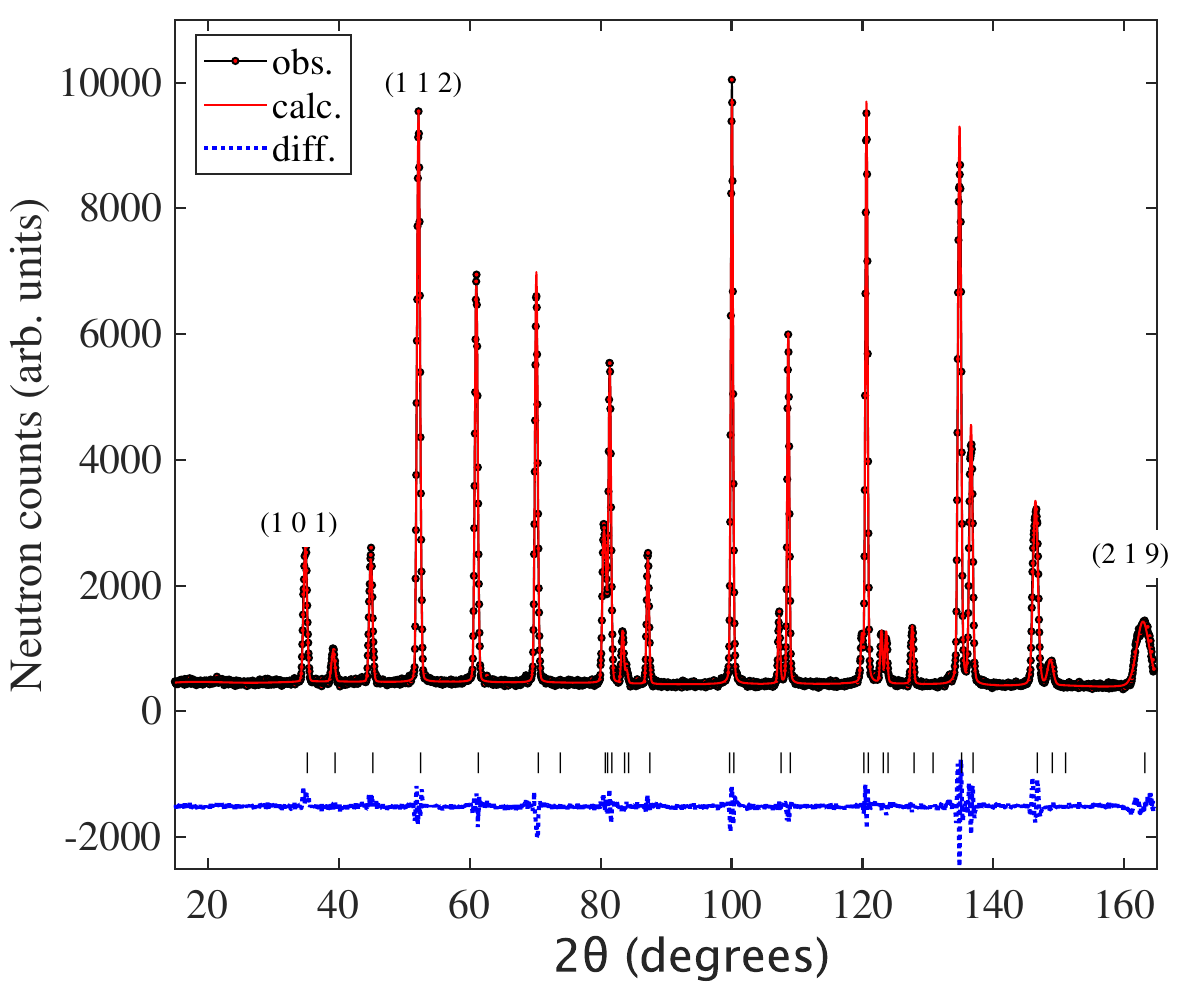} 
  \end{center}
  \caption{The Rietveld refinement pattern and the calculated profile of the neutron diffraction data for \cealge\ at T=10~K
    measured  with the wavelength $\lambda=2.45$~\AA\ at HRPT diffractometer. The rows
    of tics show the Bragg peak positions. The difference between observed and calculated intensities is shown by the dotted blue line. }
  \label{nuc_10K_2p45A_HRPT}
\end{figure}

\begin{figure}
  \begin{center}
    \includegraphics[width=\figsiz]{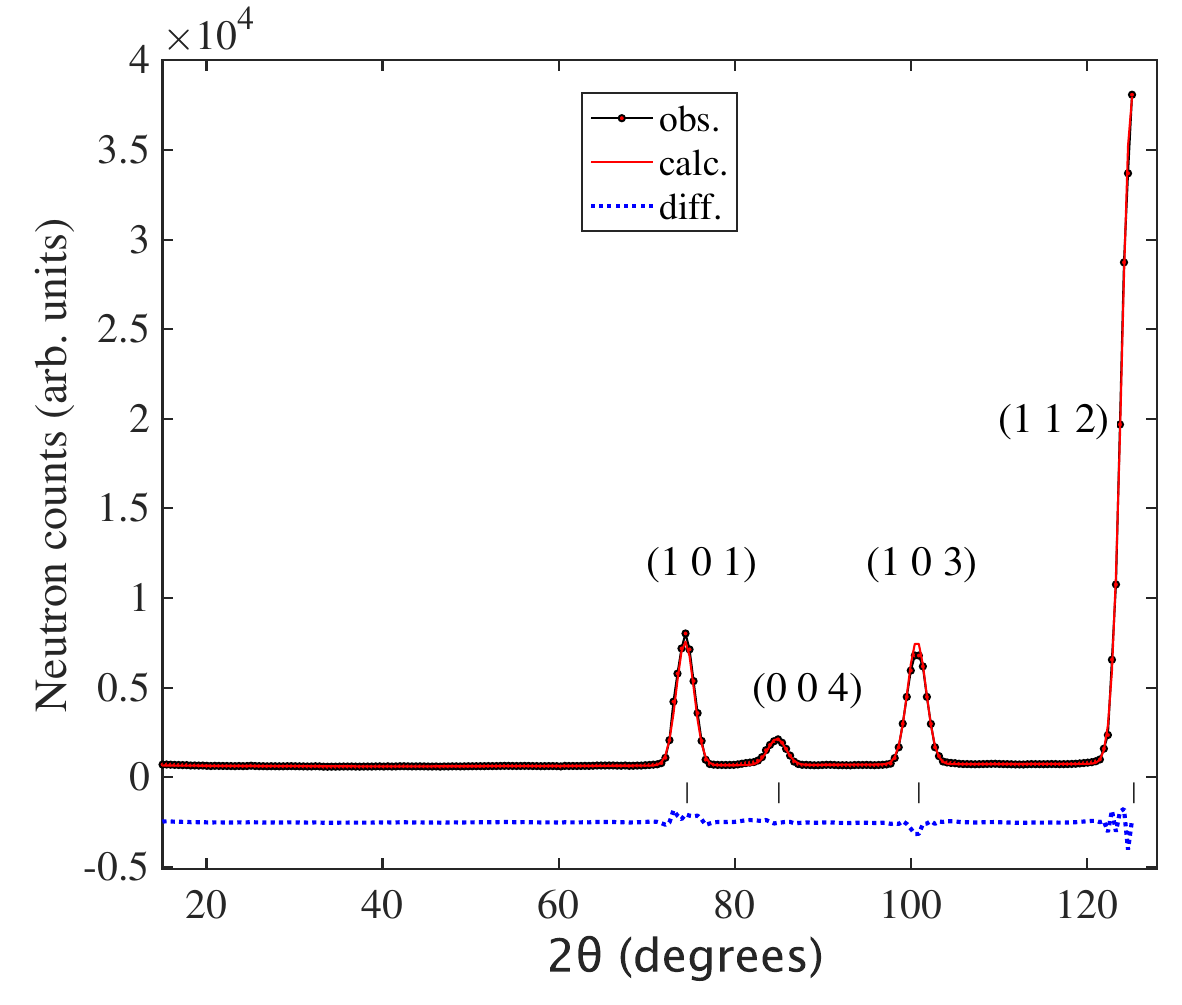} 
  \end{center}
  \caption{
  The Rietveld refinement pattern and the calculated profile of the neutron diffraction data for \cealge\ at T=10~K measured at CNCS/SNS with the wavelength $\lambda=4.96$~\AA. The rows
    of tics show the Bragg peak positions. The difference between observed and calculated intensities is shown by the dotted blue line.
}
  \label{nuc_10K_4p9A}
\end{figure}

\begin{figure}
  \begin{center}
    \includegraphics[width=\figsiz]{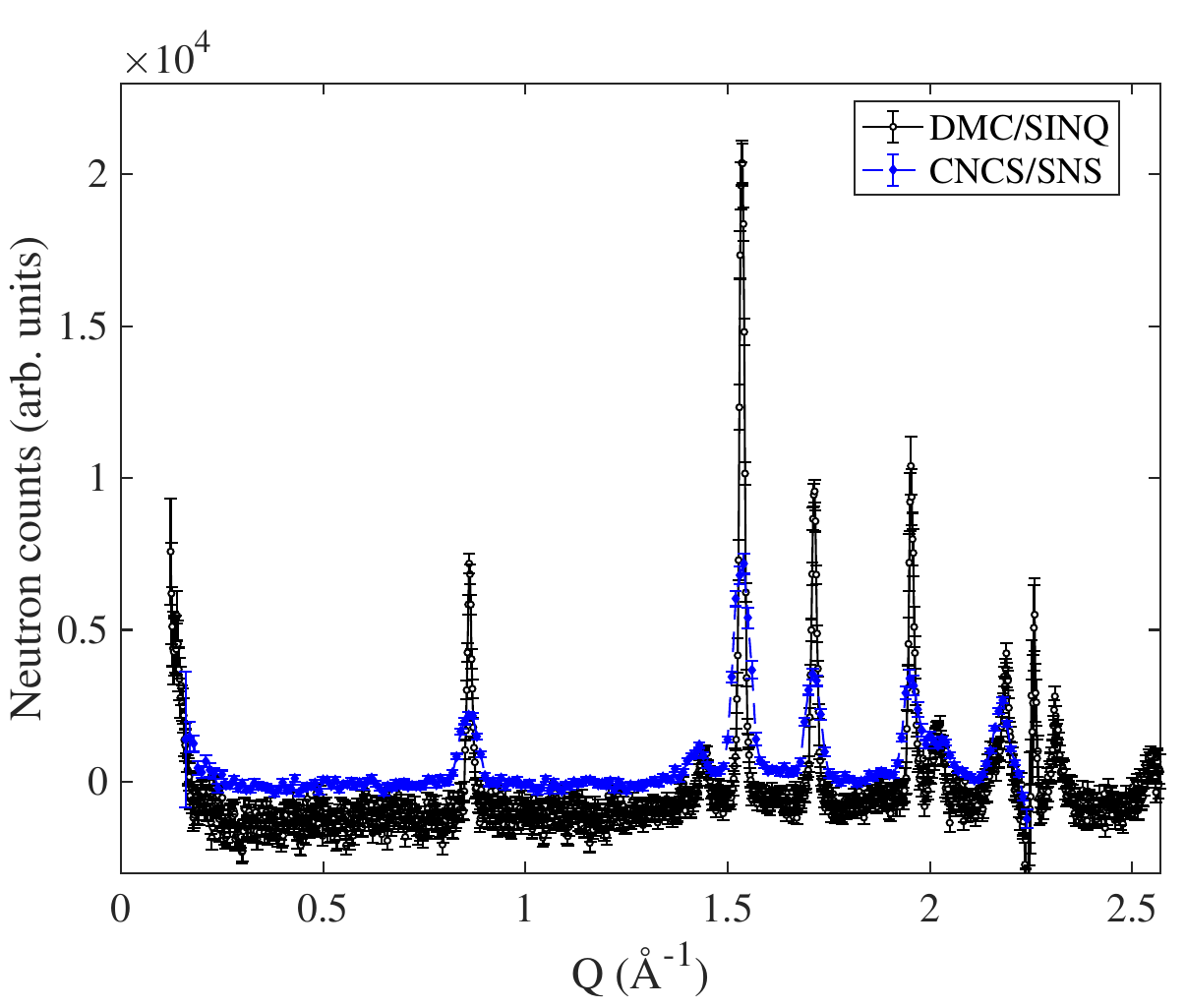} 
  \end{center}

\caption{Raw neutron diffraction difference patterns (``1.7~K - 10~K'') as a function of the scattering vector $Q$, containing purely magnetic contributions, measured at DMC/SINQ (black line and open symbols) and CNCS/SNS (blue dotted line and closed symbols) with wavelengths $\lambda = 4.506$ and 4.96~\AA, respectively. The patterns were rescaled to match the integral intensity of the diffraction peak at $Q \simeq 0.85$~\AA$^{-1}$.}

  \label{dmc_vs_cncs}
\end{figure}

\def\figsiz{8cm}

\begin{figure}
  \begin{center}

    \includegraphics[width=\figsiz]{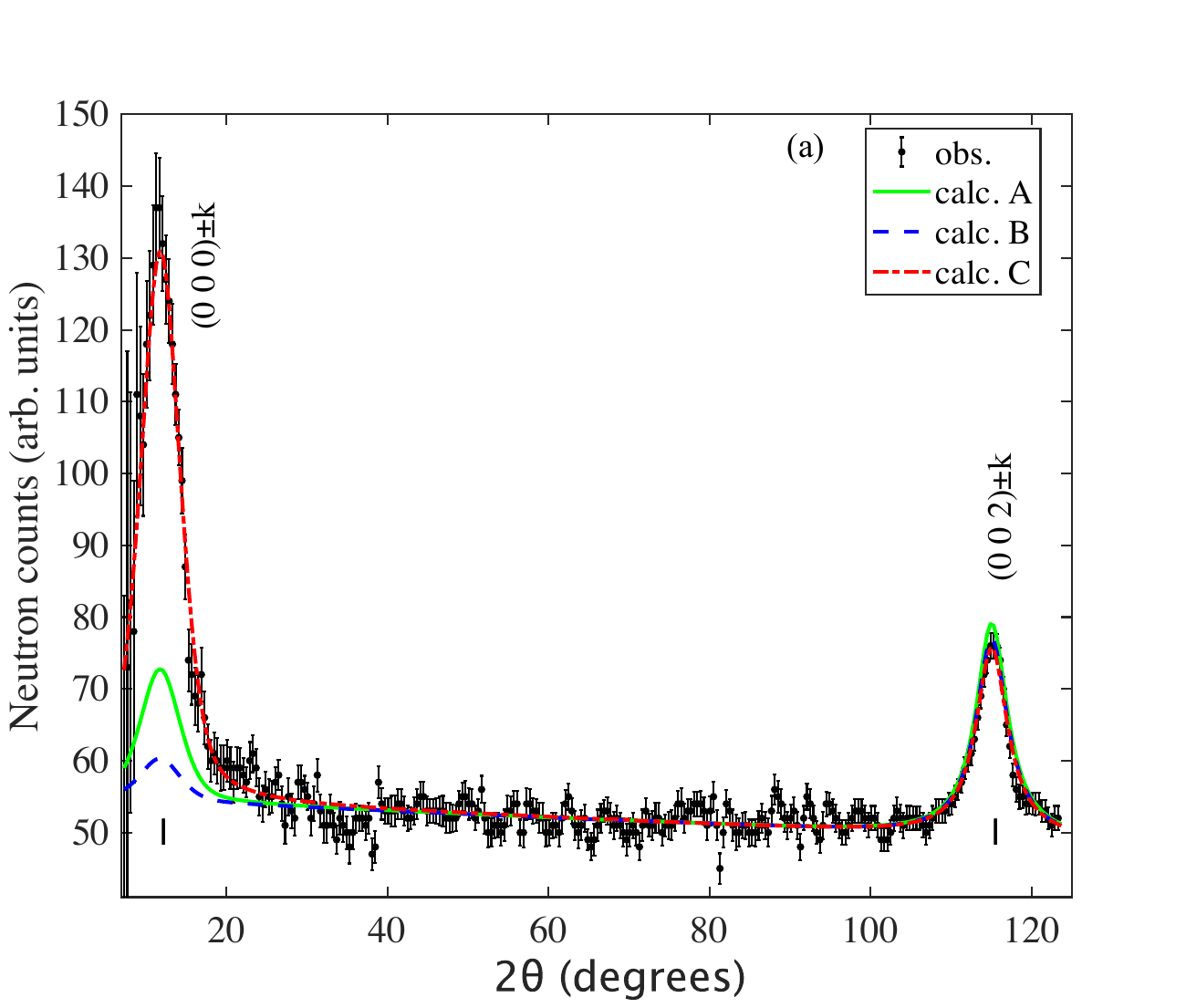} 
    \includegraphics[width=\figsiz]{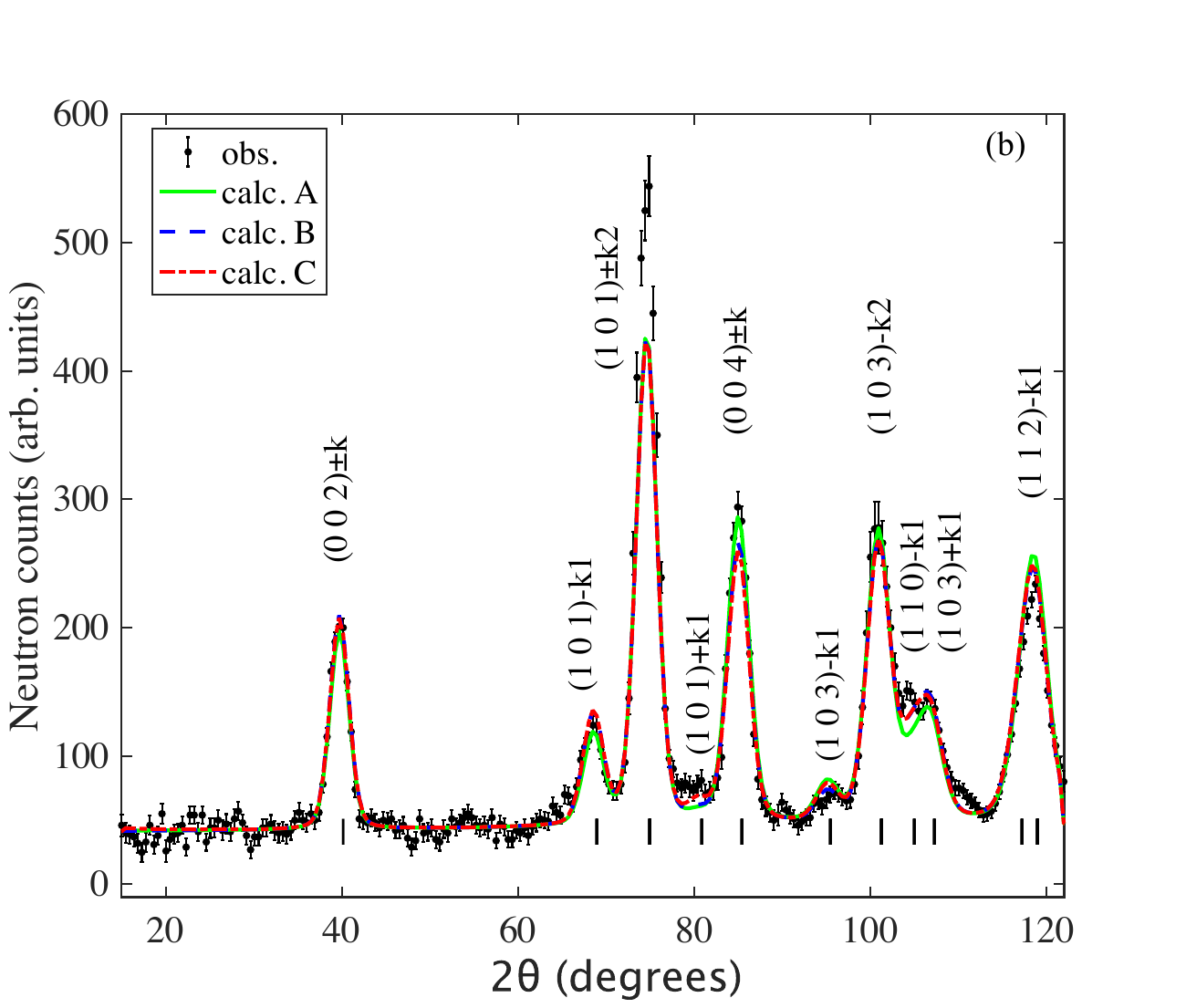} 
  \end{center}
  \caption{The difference pattern  containing purely magnetic contribution, measured at CNCS/SNS with the wavelength (a) $\lambda=12.31$~\AA\  and (b) $\lambda=4.96$~\AA. The tics show the Bragg peak positions. The calculated intensities based on model A, B and C are shown by green, red and blue dashed lines, respectively. }
  \label{dif_all_ABC}
\end{figure} 

\def\figsiz{\textwidth}

\begin{figure}
  \begin{center}
    \includegraphics[width=\figsiz]{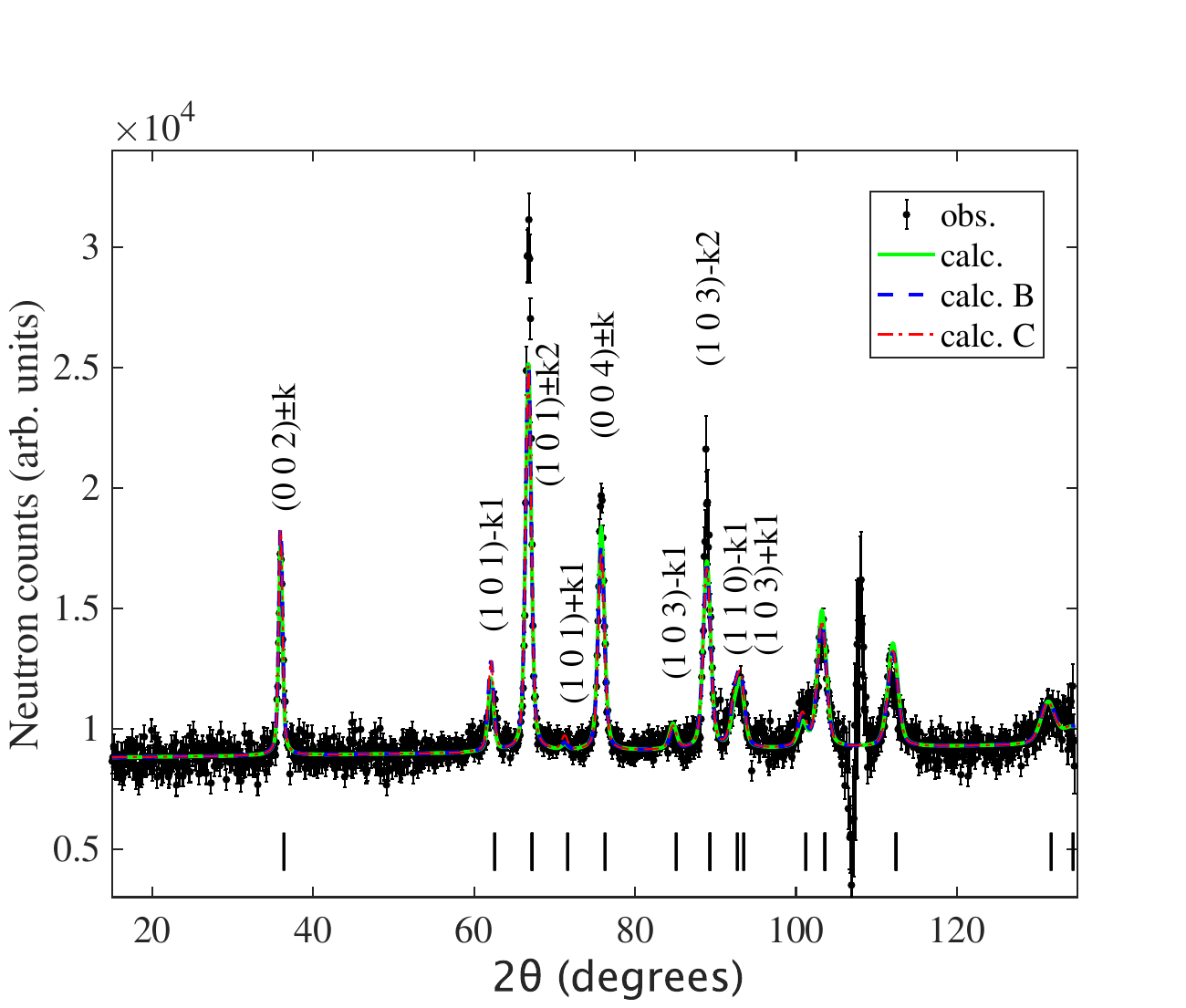} 
  \end{center}
  \caption{The difference pattern  containing purely magnetic contribution, measured at DMC with the wavelength $\lambda = 4.506$~\AA. The calculated intensities based on model A, B and C are shown by green red and blue dashed lines, respectively. }
  \label{dif_DMC_ABC}
\end{figure}

\begin{figure}
  \begin{center}
    \includegraphics[width=\figsiz]{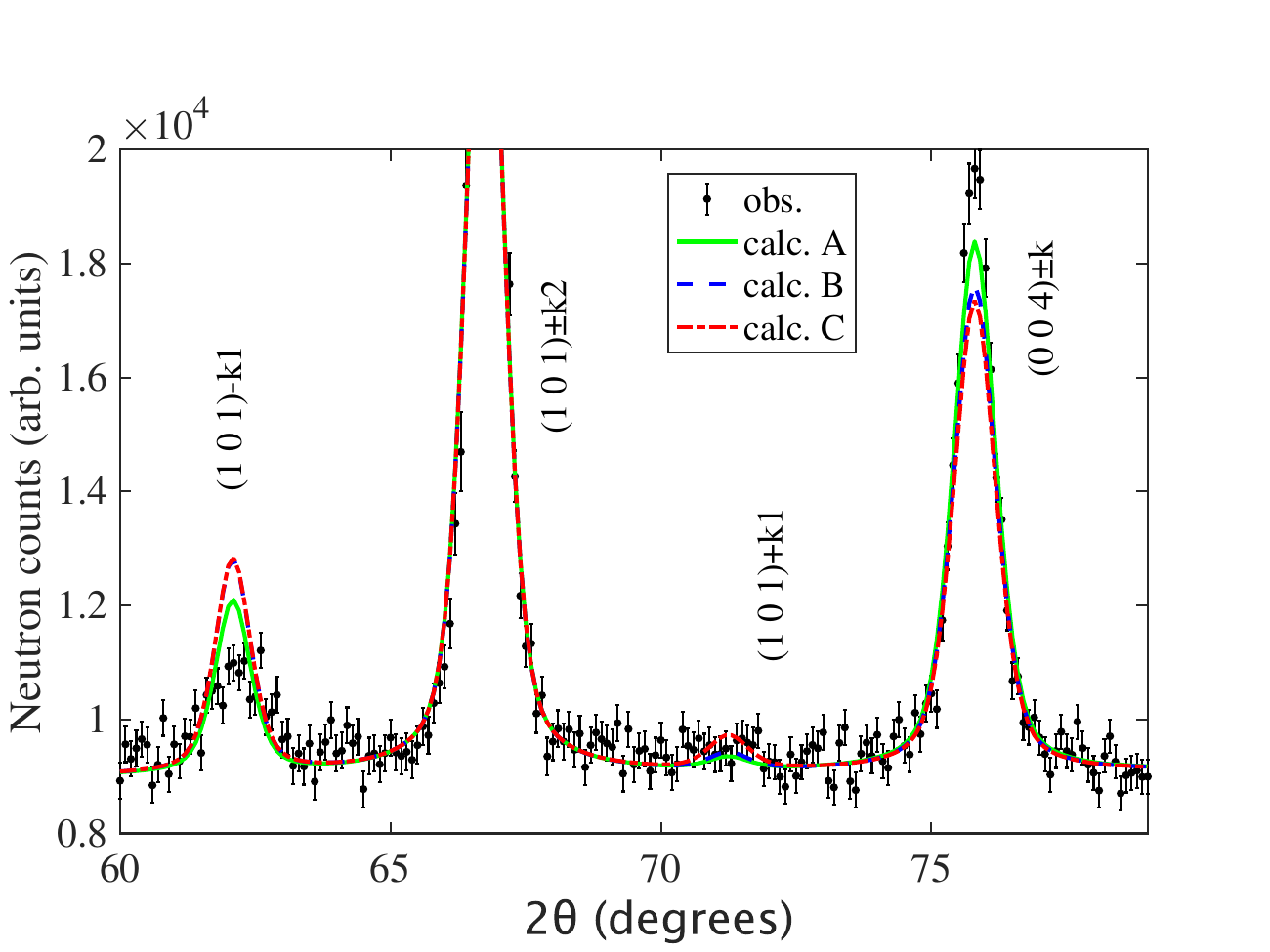} 
  \end{center}
\caption{\textcolor{blue}{Zoomed-in view of the difference pattern from Fig.\ref{dif_DMC_ABC}, highlighting subtle differences in calculated intensities among Models A, B, and C for the diffraction peaks (101)$\pm\mathbf{k}_1$ and (004)$\pm\mathbf{k}$. The full (101)$\pm\mathbf{k}_2$ peak, including its maximum, is shown in Fig.\ref{dif_DMC_ABC}.} }
  \label{dif_DMC_ABC_zoom}
\end{figure}

\def\figsiz{8cm}
\begin{figure}
  \begin{center}
    \includegraphics[width=\figsiz]{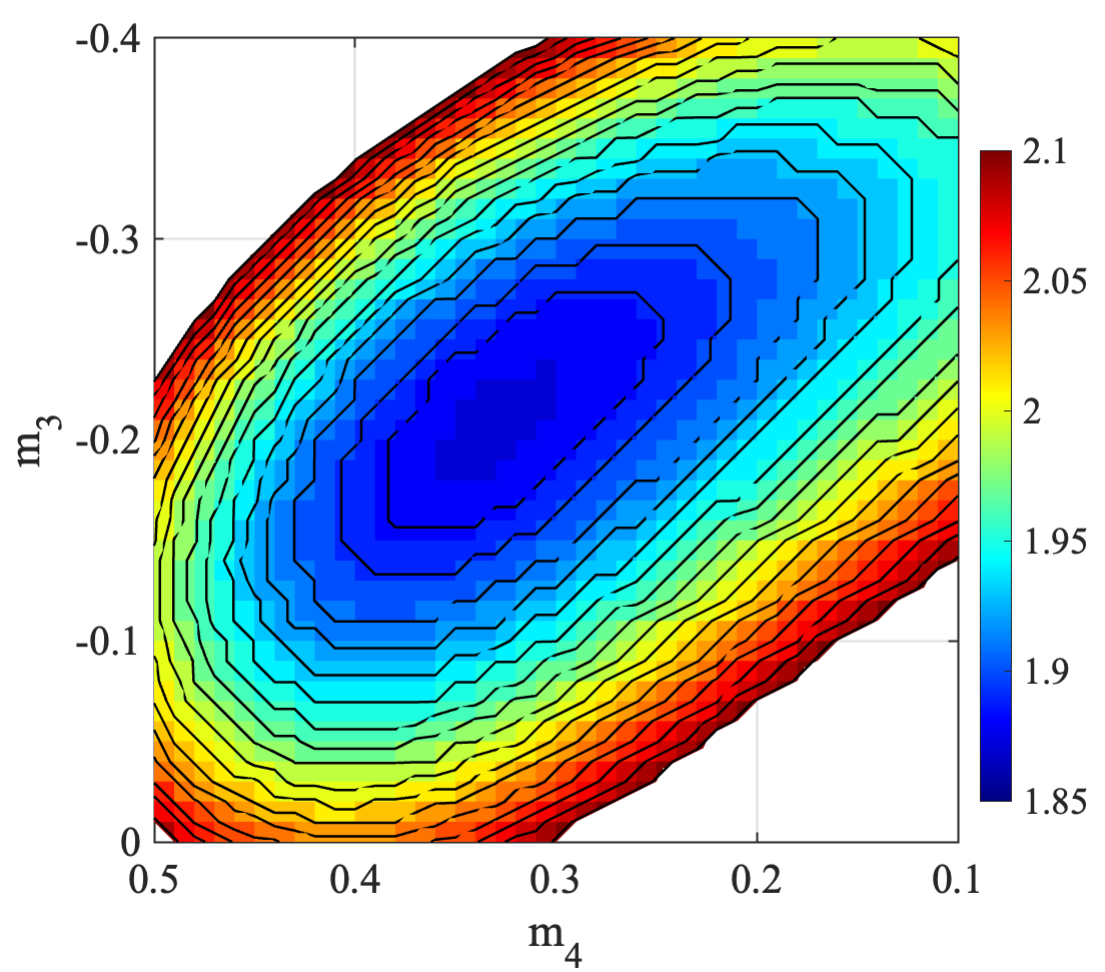} 
  \end{center}
\caption{Surface plot of $\chi^2$ as a function of the magnetic parameters $m_3$ and $m_4$ for Model A, based on refinement of the DMC dataset, with the minimum at $m_3 = -0.21(4)$, $m_4 = 0.32(5)$.}
  \label{chi2}
\end{figure}

\count1=\figsizcm

\count1=\figsizcm

\count1=\figsizcm
\multiply\count1 by 4 
\divide\count1 by 5

\def\figsiz{\textwidth}

\begin{figure}
  \begin{center}
    \includegraphics[width=\figsiz]{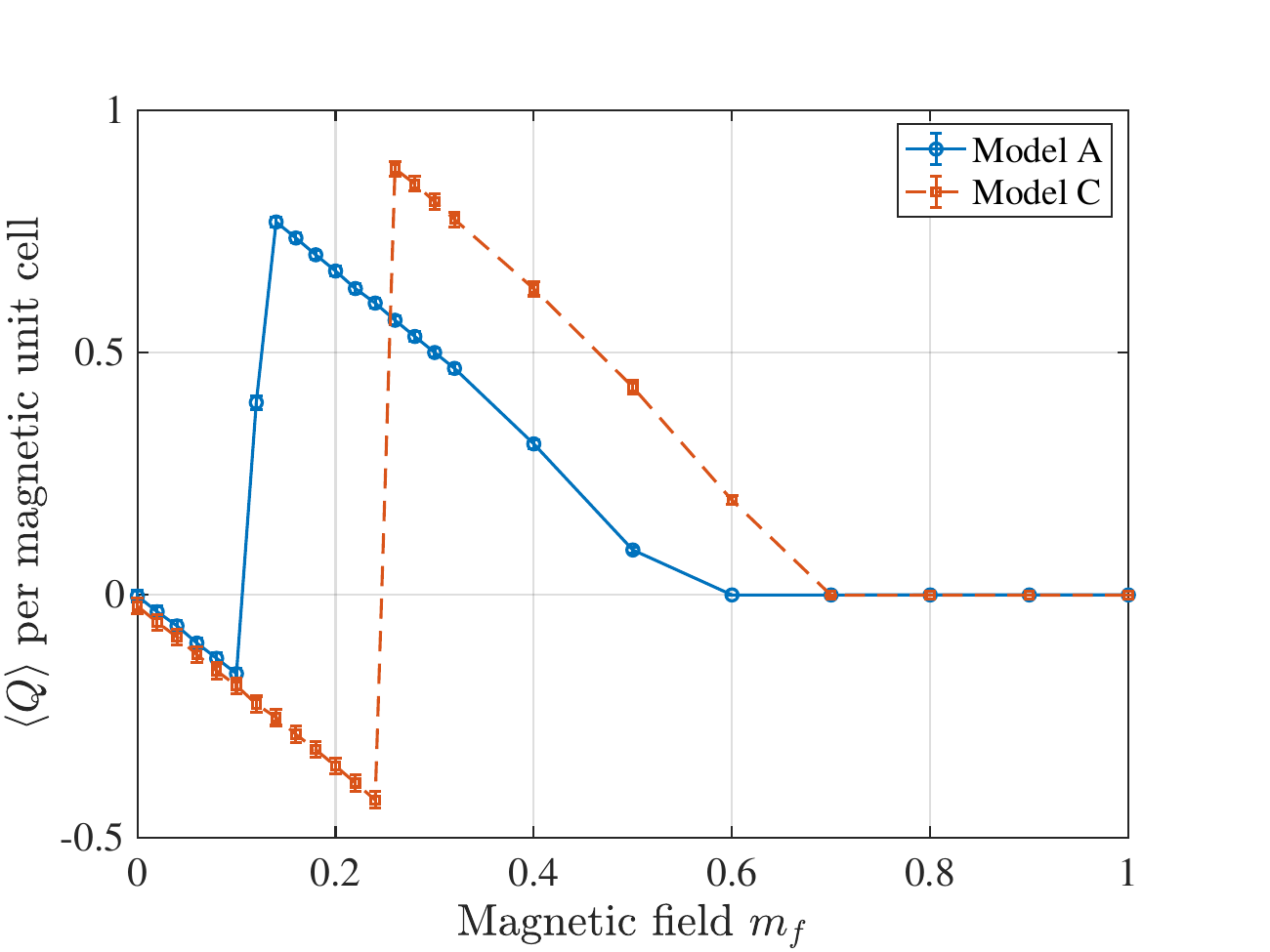} 
  \end{center}
\caption{
Average topological charge $\langle Q \rangle$ per magnetic unit cell as a function of ferromagnetic component $m_f$ along the $c$-axis for Models A and C. Error bars represent the standard error of the mean computed over the lattice. The emergence of a finite topological charge in an intermediate field range indicates the onset of a topological phase with broken meron–antimeron balance.
}
\label{Qofmf}
\end{figure}

\clearpage

\center{
\large{\bf{Supplementary materials: On the magnetic contribution of itinerant electrons to neutron diffraction in the topological antiferromagnet \cealge}}
}

\def\extgra{pdf}
\def\figsiz{\textwidth}
\def\figsiz{8cm}
\def\figsizcm{8}

\count1=\figsizcm

\begin{figure}[h]
    \centering
    \includegraphics[width=\count1cm]{Fig13a.\extgra}
    \includegraphics[width=\count1cm]{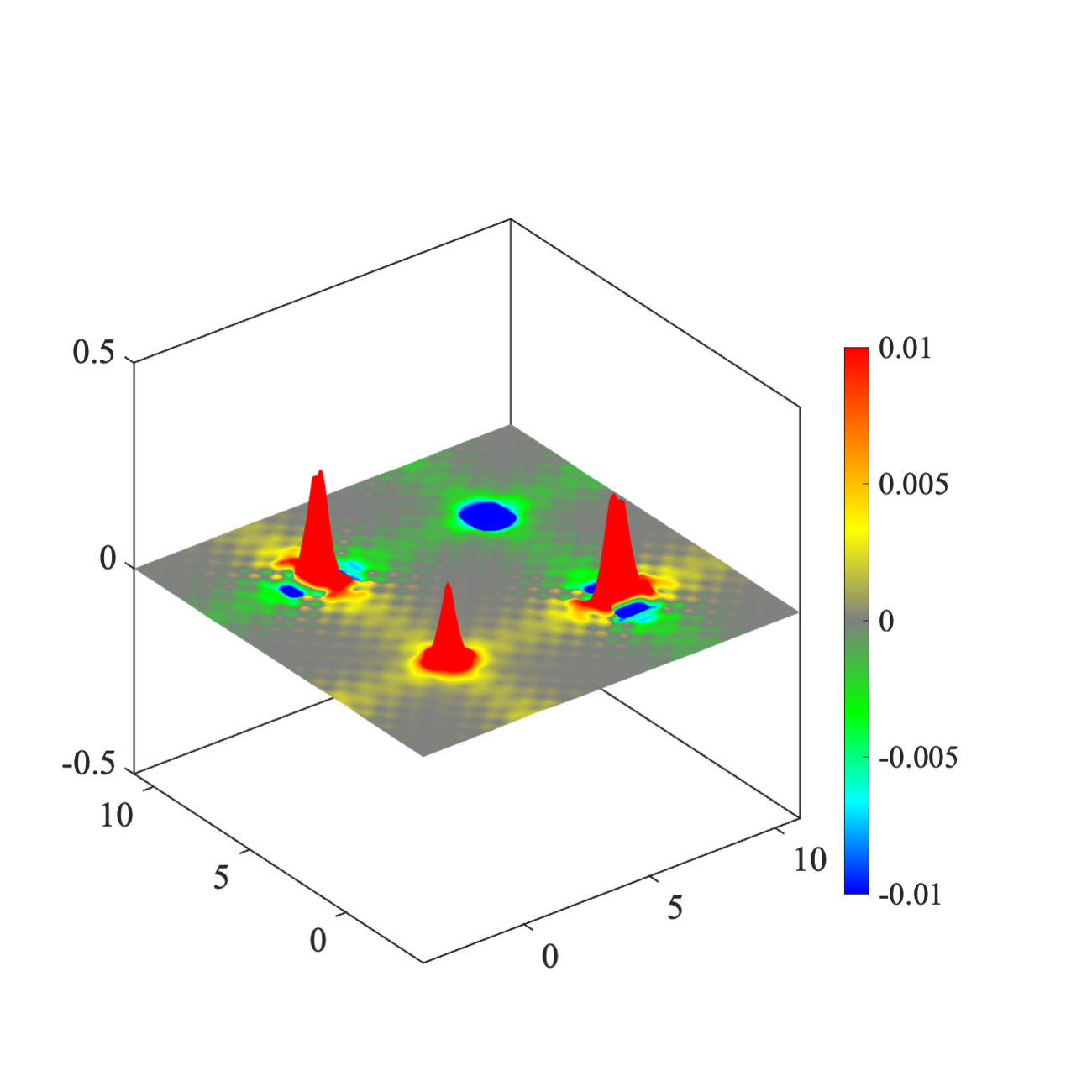}
\noindent 
Fig. SM1:
Topological charge density $w(x, y)$ calculated for Model C, illustrating its dependence on the global phase $\varphi$. The coordinates $x$ and $y$ are fractional plaquette positions in lattice units, with approximately one magnetic unit cell shown. The left panel corresponds to $\varphi = 0.0$ and the right to $\varphi = 0.12$, both with $k \simeq -0.0743$. Positive (red) and negative (blue) topological charges appear as localized features in $w(x, y)$, concentrated near the centers of each quadrant. Each quadrant carries a charge of approximately $\pm 1/2$.
\end{figure}

\count1=\figsizcm

\begin{figure}
  \begin{center}
    \includegraphics[width=16cm]{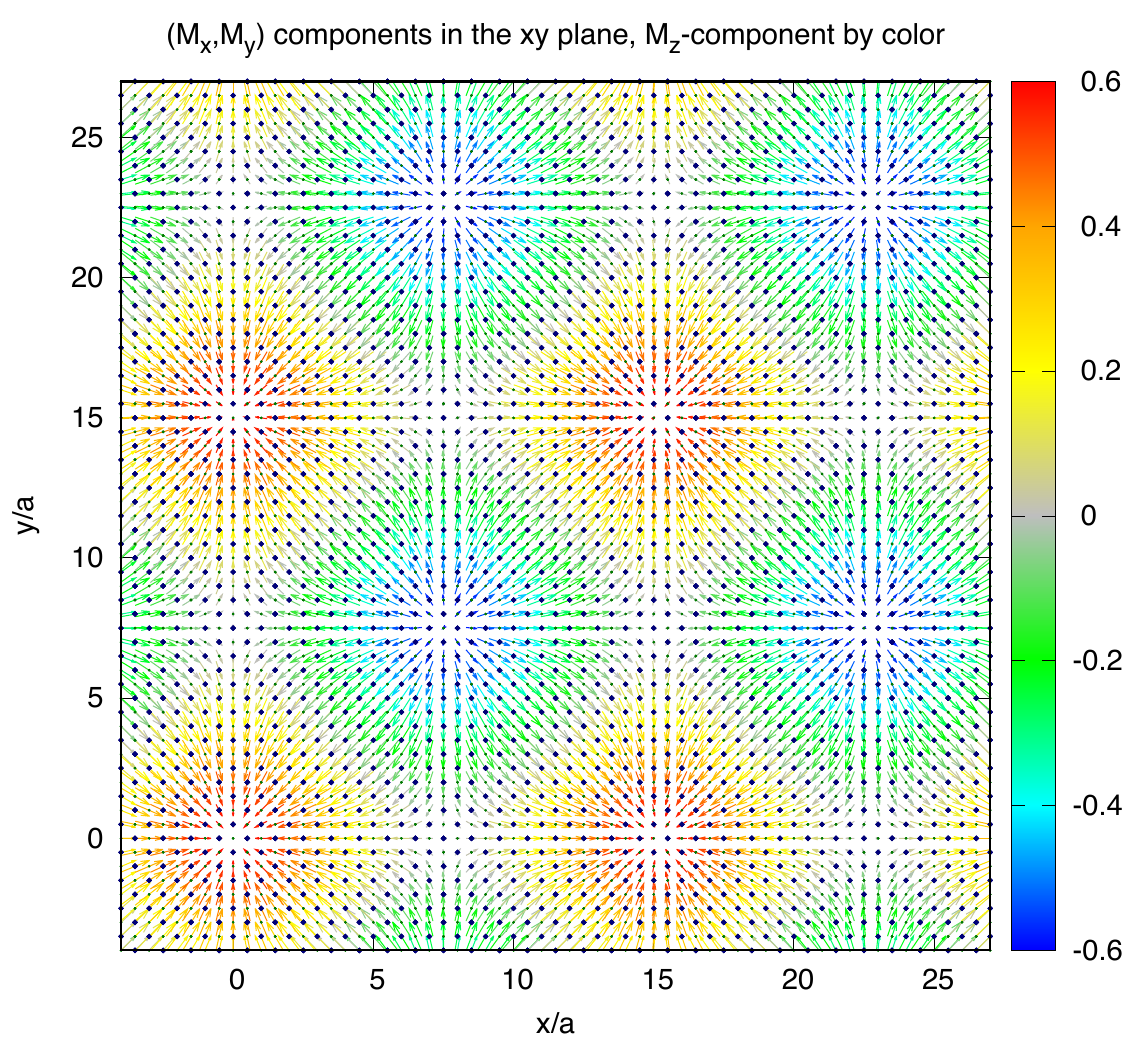}
  \end{center}
Fig. SM2:
Artificial magnetic structure of CeAlGe corresponding to a model that allows a continuous limit, with $(m_1, m_2, m_{3}, m_{4}) = (0.7, 0.7, 0.3, 0.3)$ and $k = 0.06597$, as in Model A.
The $x$- and $y$-axes are in units of the crystallographic lattice constant; the magnetic modulation period is approximately 15 unit cells.
Ce1 and Ce2 sites are shown projected onto the $xy$-plane as small green and large blue circles, respectively.
The out-of-plane ($z$) component of the magnetic moment is indicated by color.
The magnetic moments form ferromagnetic chains along the $z$-axis at each $(x,y)$ coordinate.

\end{figure}

\begin{figure}
  \begin{center}
    \includegraphics[width=\count1cm]{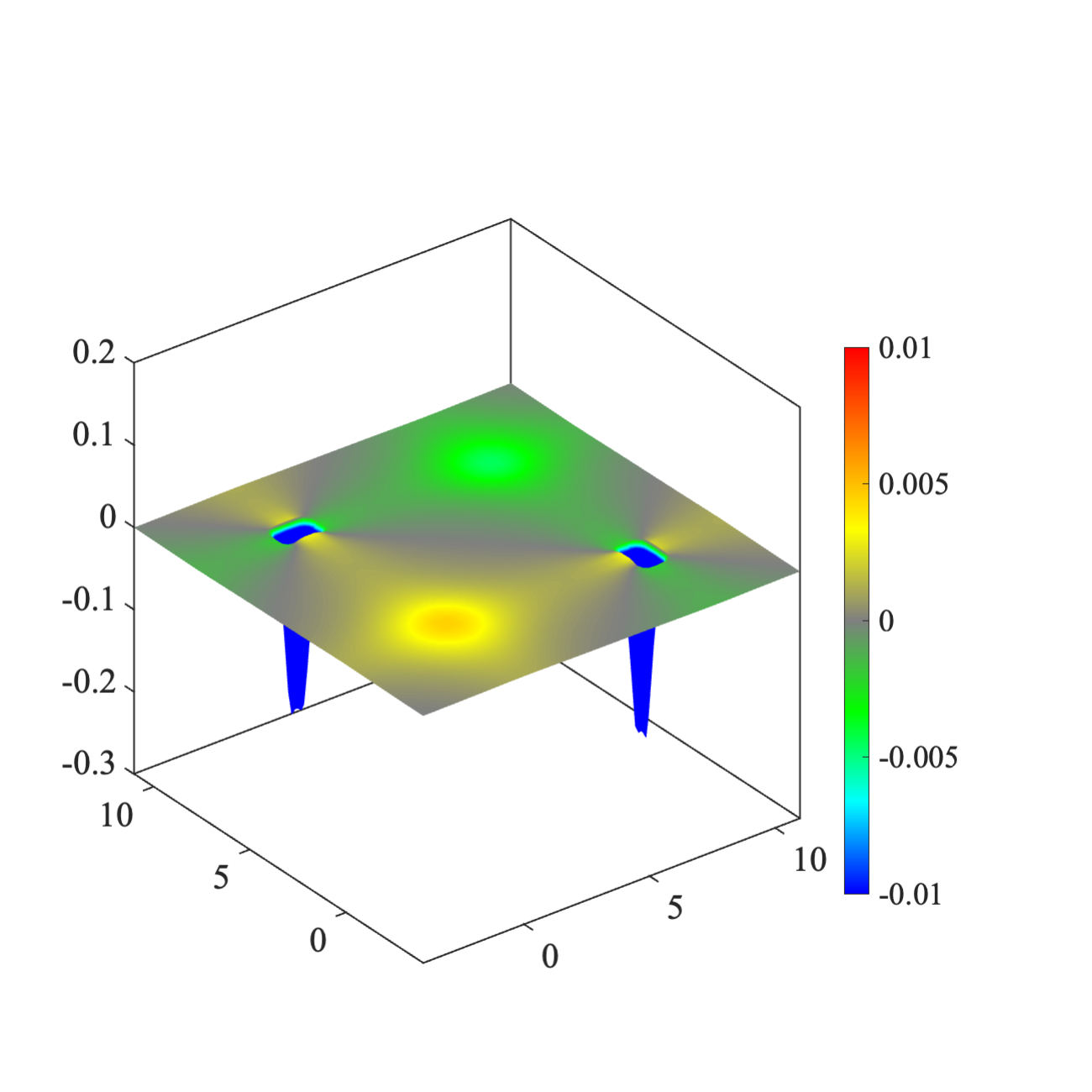}
    \includegraphics[width=\count1cm]{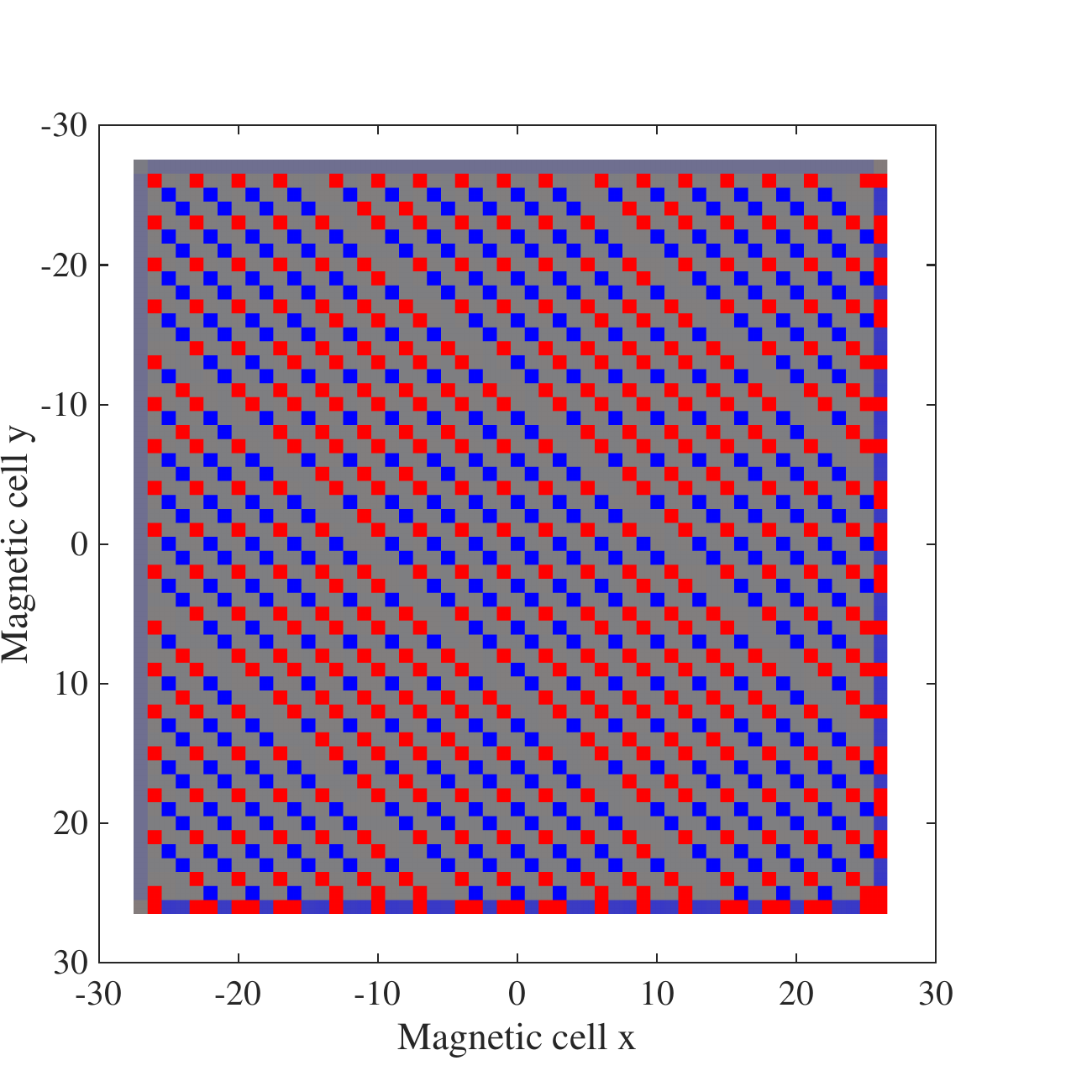}
  \end{center}
Fig. SM3:
Left: Topological charge density $w(x,y)$ calculated for the artificial magnetic structure shown in Fig.~SM2, which allows the continuous approximation. Here, $x$ and $y$ are fractional plaquette coordinates in lattice units. Approximately one magnetic unit cell is shown. Positive and negative topological charges appear as localized features in $w(x, y)$, concentrated near the quadrant centers of the magnetic unit cell. The quadrants with strong negative $w(x,y)$ carry a charge of $-1/2$, while the two other, weaker extrema carry a charge of approximately $\pm 0.3$.
Right: Topological charge $Q$ per magnetic unit cell. The axes represent integer indices $(x, y)$ of the magnetic unit cells, defined in units of the magnetic period: 15.16 crystallographic lattice units. Each magnetic cell carries a topological charge of $+1$, $0$, or $-1$, shown in red, gray, and blue, respectively. The total topological charge over the full lattice sums to zero.

\end{figure}

\renewcommand{\baselinestretch}{0.75}
\clearpage
\section*{MCIF files for magnetic models}

\subsection*{Model A}
\begin{verbatim}
#\#CIF_2.0
# Created by the Bilbao Crystallographic Server
# http://www.cryst.ehu.es
# Date: 03/12/2020

data_5yOhtAoR
_audit_creation_date            2020-03-12
_audit_creation_method          "Bilbao Crystallographic Server"






_atomic_positions_source_database_code_ICSD  .
_atomic_positions_source_other               .

_transition_temperature     4.4
_experiment_temperature     1.5

loop_
_irrep_id
_irrep_dimension
_irrep_small_dimension
_irrep_direction_type
_irrep_action
_irrep_modes_number
_irrep_presence
mSM2 4 1 special-2 primary 4 .

_exptl_crystal_magnetic_properties_details
;
NPD
;

_active_magnetic_irreps_details
;
2k incommensurate magnetic structure
2k-maximal magnetic symmetry
A 1k incommensurate structure model equally fits the data. The 2k model of
higher symmetry is prioritized in the reference.
1-dim small irrep and 4-dim full irrep along a special direction as primary
irrep, corresponding to the in-phase superposition with equal amplitudes of the 
two modulations with perpendicular k-vectors.
;

_parent_space_group.name_H-M_alt "I 4_1 m d"
_parent_space_group.IT_number             109
_parent_space_group.transform_Pp_abc  a,b,c;0,0,0

_cell_modulation_dimension               2

loop_
_cell_wave_vector_seq_id
_cell_wave_vector_x
_cell_wave_vector_y
_cell_wave_vector_z
1 0.06597 0.00000 0.00000
2 0.00000 0.06597 0.00000


_parent_space_group.child_transform_Pp_abc  a,b,c;0,0,0

_space_group.magn_ssg_name  "I4_1md1'(a,0,0)000s(0,a,0)0s0s"
_space_group.magn_point_group_name  "4mm1'"
_space_group.magn_point_group_number  13.2.45
_cell_length_a                 4.25717
_cell_length_b                 4.25717
_cell_length_c                 14.64520
_cell_angle_alpha              90.00000
_cell_angle_beta               90.00000
_cell_angle_gamma              90.00000

loop_
_space_group_symop_magn_ssg_operation.id
_space_group_symop_magn_ssg_operation.algebraic
1 x1,x2,x3,x4,x5,+1
2 -x1,-x2,x3,-x4,-x5,+1
3 -x2,x1+1/2,x3+1/4,-x5,x4,+1
4 x2,-x1+1/2,x3+1/4,x5,-x4,+1
5 -x1,x2,x3,-x4+1/2,x5+1/2,+1
6 x1,-x2,x3,x4+1/2,-x5+1/2,+1
7 x2,x1+1/2,x3+1/4,x5+1/2,x4+1/2,+1
8 -x2,-x1+1/2,x3+1/4,-x5+1/2,-x4+1/2,+1

loop_
_space_group_symop_magn_ssg_centering.id
_space_group_symop_magn_ssg_centering.algebraic
1 x1,x2,x3,x4,x5,+1
2 x1,x2,x3,x4+1/2,x5+1/2,-1
3 x1+1/2,x2+1/2,x3+1/2,x4,x5,+1
4 x1+1/2,x2+1/2,x3+1/2,x4+1/2,x5+1/2,-1

loop_
_atom_site_label
_atom_site_type_symbol
_atom_site_fract_x
_atom_site_fract_y
_atom_site_fract_z
_atom_site_occupancy
Al1 Al 0.00000 0.00000 0.18000 1
Ce1 Ce 0.00000 0.00000 0.59000 1
Ge1 Ge 0.00000 0.00000 0.01000 1


loop_
_atom_site_moment.label
_atom_site_moment.crystalaxis_x
_atom_site_moment.crystalaxis_y
_atom_site_moment.crystalaxis_z
_atom_site_moment.symmform
Ce1 0.0 0.0 0.0  0,0,0

loop_
_atom_site_Fourier_wave_vector.seq_id
_atom_site_Fourier_wave_vector.q1_coeff
_atom_site_Fourier_wave_vector.q2_coeff
1 1 0
2 0 1

loop_
_atom_site_moment_Fourier.atom_site_label
_atom_site_moment_Fourier.axis
_atom_site_moment_Fourier.wave_vector_seq_id
_atom_site_moment_Fourier_param.cos
_atom_site_moment_Fourier_param.sin
_atom_site_moment_Fourier_param.cos_symmform
_atom_site_moment_Fourier_param.sin_symmform
Ce1 x 1 0.00000 0.457(9) 0 mxs1
Ce1 y 1 0.00000 0.00000 0 0
Ce1 z 1 -0.21(4) 0.00000 mzc1 0
Ce1 x 2 0.00000 0.00000 0 0
Ce1 y 2 0.00000 1.07(1) 0 mys2
Ce1 z 2 0.33(5) 0.00000 mzc2 0
\end{verbatim}

\subsection*{Model B}
\begin{verbatim}
#\#CIF_2.0
# Created by the Bilbao Crystallographic Server
# http://www.cryst.ehu.es
# Date: 03/12/2020

data_5yOhtAoR
_audit_creation_date            2020-03-12
_audit_creation_method          "Bilbao Crystallographic Server"






_atomic_positions_source_database_code_ICSD  .
_atomic_positions_source_other               .

_transition_temperature     4.4
_experiment_temperature     1.5

loop_
_irrep_id
_irrep_dimension
_irrep_small_dimension
_irrep_direction_type
_irrep_action
_irrep_modes_number
_irrep_presence
mSM2 4 1 special-2 primary 4 .

_exptl_crystal_magnetic_properties_details
;
NPD
;

_active_magnetic_irreps_details
;
2k incommensurate magnetic structure
2k-maximal magnetic symmetry
A 1k incommensurate structure model equally fits the data. The 2k model of
higher symmetry is prioritized in the reference.
1-dim small irrep and 4-dim full irrep along a special direction as primary
irrep, corresponding to the in-phase superposition with equal amplitudes of the 
two modulations with perpendicular k-vectors.
;

_parent_space_group.name_H-M_alt "I 4_1 m d"
_parent_space_group.IT_number             109
_parent_space_group.transform_Pp_abc  a,b,c;0,0,0

_cell_modulation_dimension               2

loop_
_cell_wave_vector_seq_id
_cell_wave_vector_x
_cell_wave_vector_y
_cell_wave_vector_z
1 0.0743 0.00000 0.00000
2 0.00000 0.0743 0.00000


_parent_space_group.child_transform_Pp_abc  a,b,c;0,0,0

_space_group.magn_ssg_name  "I4_1md1'(a,0,0)000s(0,a,0)0s0s"
_space_group.magn_point_group_name  "4mm1'"
_space_group.magn_point_group_number  13.2.45
_cell_length_a                 4.25717
_cell_length_b                 4.25717
_cell_length_c                 14.64520
_cell_angle_alpha              90.00000
_cell_angle_beta               90.00000
_cell_angle_gamma              90.00000

loop_
_space_group_symop_magn_ssg_operation.id
_space_group_symop_magn_ssg_operation.algebraic
1 x1,x2,x3,x4,x5,+1
2 -x1,-x2,x3,-x4,-x5,+1
3 -x2,x1+1/2,x3+1/4,-x5,x4,+1
4 x2,-x1+1/2,x3+1/4,x5,-x4,+1
5 -x1,x2,x3,-x4+1/2,x5+1/2,+1
6 x1,-x2,x3,x4+1/2,-x5+1/2,+1
7 x2,x1+1/2,x3+1/4,x5+1/2,x4+1/2,+1
8 -x2,-x1+1/2,x3+1/4,-x5+1/2,-x4+1/2,+1

loop_
_space_group_symop_magn_ssg_centering.id
_space_group_symop_magn_ssg_centering.algebraic
1 x1,x2,x3,x4,x5,+1
2 x1,x2,x3,x4+1/2,x5+1/2,-1
3 x1+1/2,x2+1/2,x3+1/2,x4,x5,+1
4 x1+1/2,x2+1/2,x3+1/2,x4+1/2,x5+1/2,-1

loop_
_atom_site_label
_atom_site_type_symbol
_atom_site_fract_x
_atom_site_fract_y
_atom_site_fract_z
_atom_site_occupancy
Al1 Al 0.00000 0.00000 0.18000 1
Ce1 Ce 0.00000 0.00000 0.59000 1
Ge1 Ge 0.00000 0.00000 0.01000 1


loop_
_atom_site_moment.label
_atom_site_moment.crystalaxis_x
_atom_site_moment.crystalaxis_y
_atom_site_moment.crystalaxis_z
_atom_site_moment.symmform
Ce1 0.0 0.0 0.0  0,0,0

loop_
_atom_site_Fourier_wave_vector.seq_id
_atom_site_Fourier_wave_vector.q1_coeff
_atom_site_Fourier_wave_vector.q2_coeff
1 1 0
2 0 1

loop_
_atom_site_moment_Fourier.atom_site_label
_atom_site_moment_Fourier.axis
_atom_site_moment_Fourier.wave_vector_seq_id
_atom_site_moment_Fourier_param.cos
_atom_site_moment_Fourier_param.sin
_atom_site_moment_Fourier_param.cos_symmform
_atom_site_moment_Fourier_param.sin_symmform
Ce1 x 1 0.00000 0.42(1) 0 mxs1
Ce1 y 1 0.00000 0.00000 0 0
Ce1 z 1 -0.30(5) 0.00000 mzc1 0
Ce1 x 2 0.00000 0.00000 0 0
Ce1 y 2 0.00000 1.06(1) 0 mys2
Ce1 z 2 0.37(7) 0.00000 mzc2 0
\end{verbatim}

\subsection*{Model C}
\begin{verbatim}
#\#CIF_2.0
# Created by the Bilbao Crystallographic Server
# http://www.cryst.ehu.es
# Date: 03/12/2020

data_5yOhtAoR
_audit_creation_date            2020-03-12
_audit_creation_method          "Bilbao Crystallographic Server"






_atomic_positions_source_database_code_ICSD  .
_atomic_positions_source_other               .

_transition_temperature     4.4
_experiment_temperature     1.5

loop_
_irrep_id
_irrep_dimension
_irrep_small_dimension
_irrep_direction_type
_irrep_action
_irrep_modes_number
_irrep_presence
mSM2 4 1 special-2 primary 4 .

_exptl_crystal_magnetic_properties_details
;
NPD
;

_active_magnetic_irreps_details
;
2k incommensurate magnetic structure
2k-maximal magnetic symmetry
A 1k incommensurate structure model equally fits the data. The 2k model of
higher symmetry is prioritized in the reference.
1-dim small irrep and 4-dim full irrep along a special direction as primary
irrep, corresponding to the in-phase superposition with equal amplitudes of the 
two modulations with perpendicular k-vectors.
;

_parent_space_group.name_H-M_alt "I 4_1 m d"
_parent_space_group.IT_number             109
_parent_space_group.transform_Pp_abc  a,b,c;0,0,0

_cell_modulation_dimension               2

loop_
_cell_wave_vector_seq_id
_cell_wave_vector_x
_cell_wave_vector_y
_cell_wave_vector_z
1 0.0743 0.00000 0.00000
2 0.00000 0.0743 0.00000


_parent_space_group.child_transform_Pp_abc  a,b,c;0,0,0

_space_group.magn_ssg_name  "I4_1md1'(a,0,0)000s(0,a,0)0s0s"
_space_group.magn_point_group_name  "4mm1'"
_space_group.magn_point_group_number  13.2.45
_cell_length_a                 4.25717
_cell_length_b                 4.25717
_cell_length_c                 14.64520
_cell_angle_alpha              90.00000
_cell_angle_beta               90.00000
_cell_angle_gamma              90.00000

loop_
_space_group_symop_magn_ssg_operation.id
_space_group_symop_magn_ssg_operation.algebraic
1 x1,x2,x3,x4,x5,+1
2 -x1,-x2,x3,-x4,-x5,+1
3 -x2,x1+1/2,x3+1/4,-x5,x4,+1
4 x2,-x1+1/2,x3+1/4,x5,-x4,+1
5 -x1,x2,x3,-x4+1/2,x5+1/2,+1
6 x1,-x2,x3,x4+1/2,-x5+1/2,+1
7 x2,x1+1/2,x3+1/4,x5+1/2,x4+1/2,+1
8 -x2,-x1+1/2,x3+1/4,-x5+1/2,-x4+1/2,+1

loop_
_space_group_symop_magn_ssg_centering.id
_space_group_symop_magn_ssg_centering.algebraic
1 x1,x2,x3,x4,x5,+1
2 x1,x2,x3,x4+1/2,x5+1/2,-1
3 x1+1/2,x2+1/2,x3+1/2,x4,x5,+1
4 x1+1/2,x2+1/2,x3+1/2,x4+1/2,x5+1/2,-1

loop_
_atom_site_label
_atom_site_type_symbol
_atom_site_fract_x
_atom_site_fract_y
_atom_site_fract_z
_atom_site_occupancy
Al1 Al 0.00000 0.00000 0.18000 1
Ce1 Ce 0.00000 0.00000 0.59000 1
Ge1 Ge 0.00000 0.00000 0.01000 1


loop_
_atom_site_moment.label
_atom_site_moment.crystalaxis_x
_atom_site_moment.crystalaxis_y
_atom_site_moment.crystalaxis_z
_atom_site_moment.symmform
Ce1 0.0 0.0 0.0  0,0,0

loop_
_atom_site_Fourier_wave_vector.seq_id
_atom_site_Fourier_wave_vector.q1_coeff
_atom_site_Fourier_wave_vector.q2_coeff
1 1 0
2 0 1

loop_
_atom_site_moment_Fourier.atom_site_label
_atom_site_moment_Fourier.axis
_atom_site_moment_Fourier.wave_vector_seq_id
_atom_site_moment_Fourier_param.cos
_atom_site_moment_Fourier_param.sin
_atom_site_moment_Fourier_param.cos_symmform
_atom_site_moment_Fourier_param.sin_symmform
Ce1 x 1 0.00000 0.41(1) 0 mxs1
Ce1 y 1 0.00000 0.00000 0 0
Ce1 z 1 -0.21(1) 0.00000 mzc1 0
Ce1 x 2 0.00000 0.00000 0 0
Ce1 y 2 0.00000 1.04(1) 0 mys2
Ce1 z 2 0.47(1) 0.00000 mzc2 0
\end{verbatim}

\end{document}